\documentclass[twocolumn,           
               showpacs,            
               nopreprintnumbers,     
               aps,                 
               prd,          	    
               letterpaper,             
              groupeaddress,      
               nofootinbib,         
               tightenlines,        
               floats,floatfix,      
               showkeys
               ]{revtex4-1}
               
\usepackage[toc,page]{appendix}
\usepackage{graphicx}
\usepackage{dcolumn}
\usepackage{bm}
\usepackage{amsmath}
\usepackage{amsfonts,amssymb}
\usepackage{soul}
\usepackage{color}
\definecolor{v}{rgb}{0.6, 0.2, 0.8} 
\definecolor{MAGA}{rgb}{0.1, 0.43, 0.75}
\definecolor{jm}{rgb}{0.13, 0.48, 0.64}

\usepackage{xcolor}
\usepackage{enumerate}
\usepackage{float}
\usepackage{subfig}
\usepackage{ulem}

\begin{document}

\title{The Universe acceleration from the Unimodular gravity view point: Background and linear perturbations}

\author{Miguel A. Garc\'ia-Aspeitia$^{1,2}$}
\email{aspeitia@fisica.uaz.edu.mx}

\author{A. Hern\'andez-Almada$^3$}

\author{Juan Maga\~na$^{4}$}

\author{V. Motta$^5$}

\affiliation{$^1$Unidad Acad\'emica de F\'isica, Universidad Aut\'onoma de Zacatecas, Calzada Solidaridad esquina con Paseo a la Bufa S/N C.P. 98060, Zacatecas, M\'exico.}
\affiliation{$^2$Consejo Nacional de Ciencia y Tecnolog\'ia, \\ Av. Insurgentes Sur 1582. Colonia Cr\'edito Constructor, Del. Benito Ju\'arez C.P. 03940, Ciudad de M\'exico, M\'exico.}
\affiliation{$^3$Facultad de Ingenier\'ia, Universidad Aut\'onoma de Quer\'etaro, Centro Universitario Cerro de las Campanas, 76010, Santiago de Quer\'etaro, M\'exico.}
\affiliation{$^4$Instituto de Astrof\'isica \& Centro de Astro-Ingenier\'ia, Pontificia Universidad Cat\'olica de Chile, \\Av. Vicu\~na Mackenna, 4860, Santiago, Chile}
\affiliation{$^5$Instituto de F\'isica y Astronom\'ia, Facultad de Ciencias, Universidad de Valpara\'iso, Avda. Gran Breta\~na 1111, Valpara\'iso, Chile.}

\begin{abstract}
With the goal of studying the cosmological constant (CC) problem, we present an exhaustive analysis of unimodular gravity as a possible candidate to resolve the CC origin and with this, the current Universe acceleration. In this theory, a correction constant (CC-like) in the field equations sources the late cosmic acceleration. This constant is related to a new parameter, $z_{ini}$, which is interpreted as the redshift of CC-like emergence. By comparing with the CC value obtained from Planck and Supernovaes measurements, it is possible to estimate $z_{ini}=11.15^{+0.01}_{-0.02}$ and $z_{ini}=11.43^{+0.03}_{-0.06}$ respectively, which is close to the reionization epoch. Moreover, we use the observational Hubble data (OHD), Type Ia Supernovae (SnIa), Baryon Acoustic Oscillations (BAO) and the Cosmic Microwave Background Radiation (CMB) distance data to constrain the UG cosmological parameters. A Joint analysis (OHD+SnIa+BAO+CMB), results in $z_{ini}=11.47^{+0.074}_{-0.073}$ within $1\sigma$ confidence level consistent with our estimation from Planck and Supernovae measurements.
We also include linear perturbations, starting with scalar and tensor perturbations and complementing with the perturbed Boltzmann equation for photons. We show that the $00$ term in the UG field equations and the Boltzmann equation for photons contains corrections, meanwhile the other equations are similar as those obtained in standard cosmology.
\end{abstract}

\date{\today}
\maketitle

\section{Introduction}

Universe acceleration is one of the most intriguing conundrums in modern cosmology \cite{Riess:1998,Perlmutter,Scolnic:2017caz}. 
This feature has been confirmed not only by Supernovae of the Type Ia (SnIa), but also with observations related to Cosmic Microwave Background (CMB) Radiation \cite{Planck:2018}, among others\footnote{See other recent compilations of observations like Cosmic Chronometers \cite{Magana:2017nfs} and Strong Gravitational Lensing (SGL) \cite{Amante:2019xao} to test theoretical models related with the Universe acceleration.}. Nowadays there is a concentration of efforts to understand the physics (at macro and micro scales) responsible for such acceleration (see \cite{Copeland:2006wr,Li:2011sd} for a compilation of models) and its consequences in the Universe evolution.
Despite that many hypotheses have been put on the table, the best candidate is still the cosmological constant (CC), an essential ingredient in the $\Lambda$CDM model representing the $\sim69\%$ of the total components in the Universe \cite{Planck:2018}. 
One feature of CC is that it is imposed in the Einstein field equations, sustained only by the Lovelock's theorem \cite{Lovelock}, and its energy density value is adjusted by cosmological observations to reproduce the expected Universe dynamics. Nevertheless, the physics of the CC is still unknown and the preferred interpretation is that CC is associated with the quantum vacuum fluctuation of the space-time due to its auto-gravitating characteristics \cite{Weinberg,Zeldovich}. This hypothesis leads to one of the biggest discrepancies between theory and observation in modern physics, because the theoretical prediction according to quantum field theory is $\sim 120$ orders of magnitude greater than the obtained from cosmological observations \cite{Weinberg,Zeldovich}.

On the other hand, unimodular gravity (UG) proposes an interesting and natural origin of the CC \cite{James,Ellis,Gao:2014nia,Josset:2016,Perez:2018wlo,Perez:2017krv}, since it appears as an integration constant, which can be chosen without extra hypothesis and estimated with cosmological observations. This particular permissiveness comes from the consideration that the determinant of the metric is a constant invariant volume form, reducing the degrees of freedom via diffeomorphism invariance in the Einstein-Hilbert action and allowing a non gravitating CC. Using a constrained Hamiltonian dynamic, it is possible to show that UG has the same number of degrees of freedom than General Relativity (GR), even though it has less symmetries, and thus avoiding the problems that other theories encounter when working with third order derivatives \cite{Henn}. In addition, UG does not generate a natural energy-momentum conservation and its assumption should be imposed as an extra hypothesis, reducing it automatically to the traditional Einstein field equations plus the integration constant discussed above \cite{Gao:2014nia}. However, from our point of view, such extra hypothesis reduces important characteristics of UG, making blurry some aspects that could lead us to a most profound theory of the space-time itself while maintaining the CC in the same mystery as in the standard paradigm. 

A radical point of view is to disregard the traditional hypothesis of energy-momentum conservation (as previous works do \cite{Ellis,Gao:2014nia}), allowing a new mathematical structure of conservation which not only contains the energy-momentum tensor but also the Ricci and energy-momentum scalar, and whose structure emerges naturally from the UG field equations without extra hypothesis. This new standpoint gives a mechanism to understand the CC from a different perspective, resolving the value of the energy density and tracing its presence as far as the epoch of electro-weak transition \cite{Perez:2018wlo,Perez:2017krv}. Indeed, in this approach the CC can be interpreted as a cumulative small violation of the energy-momentum conservation, which in turn should be understood as granularity of the space-time structure \cite{Perez:2017krv}, being a possible evidence of a quantum space-time.

In addition, UG could be the candidate to extend GR because is compatible with observations of gravitational waves \cite{Abbott:2016blz,Abbott:2016nmj} or super massive black holes \cite{MassiveBlackHoles} and could be the key to explore the quantum regime  \cite{Bufalo:2015wda,Percacci:2017fsy} or even the quantum gravity realm \cite{Perez:2018wlo,Perez:2017krv}. As final comments, the UG theory has not only been tested in cosmology but also in compact objects to constrain the violation of the energy-momentum conservation and its consequences in astrophysics \cite{Astorga-Moreno:2019}.

In this vein, in Ref. \cite{Garcia-Aspeitia:2019yni} the authors describe the CC in the unimodular gravity context, obtaining a constant parameter  related to the equation of state of radiation. This new radical approach suggests that the causative of Universe acceleration is a constant related to the violation of the energy-momentum conservation at a given component, such as radiation, which emerges at a redshift $z_{ini}$, close the \textit{reionization epoch}. Another characteristic is that, unlike GR, unimodular gravity can provide an Universe acceleration without invoking an exotic fluid with a negative equation of state (EoS), which is an important advantage in comparison with GR. In addition, reference \cite{Garcia-Aspeitia:2019yni} suggests that the physics at early times could be different from the physics at late times due to the emergence of this new constant term 
that later will act like an effective CC  (hereafter CC-like). Hence, this constant term in Friedmann equations should be the smoking gun of the model that could be detected in future experiments in order to refute or validate it. In this paper, we constrain the free parameters of the model through diverse observation (OHD, SnIa, CMB and BAO data) concluding that the epoch of CC-like emergence should be in the reionization epoch as pointed out (based in CMB data) by Ref. \cite{Garcia-Aspeitia:2019yni}, contributing with observational evidence of these assertions. Even demonstrating that there is no tension between the different data samples strengthening the presented model. Moreover, we discuss diverse phenomena that could lead to understand the emergence of the CC-like, such as those related to the continuous spontaneous localization. Finally, this particular scenario of \textit{emergent} models (see for example \cite{Hernandez-Ganly:2020}) could give us clues about the tension between the Hubble constant values estimated from the early and the late Universe data \cite{Verde:2019ivm} (see also \cite{Perez:2020cwa,DiValentino:2021izs,Efstathiou:2021ocp}).

On the other hand, linear perturbations in UG are required to study those effects related to the growth of structure, the CMB power spectrum, among others, and to develop a comparison with the standard perturbations that rise from standard cosmology under GR context. Recently, in \cite{Basak:2015}, the authors studied linear perturbations with only a scalar field concluding that UG and GR are similar at this level. However, the result is centered in adding the hypothesis that $\nabla^{\mu}[(R+8\pi GT)g_{\mu\nu}]=0$, which is an extra assumption and directly converge to GR with subtle changes. Here, we develop linear perturbations in the Newtonian gauge line element to scalar and tensor perturbations, showing that the $00$ component contains an extra term that involves the perturbations for Ricci and energy-momentum scalars, meanwhile the $ij$ component and tensor perturbations are identically to those predicted by GR. In contrast, the Boltzmann equation for photons is exposed because it contains the energy momentum violations that characterize the UG. Notoriously, the extra term carries higher order derivatives in the conformal time component for the scale factor and for the scalar curvature. If we add the extra hypothesis that the derivative of the extra term in UG is equal to zero, we recover the standard linear equations for GR. A new jerk function must be constructed in order to follow the paths of the homogeneous and isotropic scenario previously discussed and help us to discern among spurious solutions caused by third order derivatives.

The present paper is organized as follows: In Sec. \ref{Sec:Review} we present a review of UG based in previous studies. In Sec. 
\ref{cosmological} we constrain UG free parameters through OHD, SnIa, CMB posterior distances and BAO samples; containing at the end the respectively results of this section. In Sec. \ref{PERT} we present the equations associated to linear perturbations, dividing our study in scalar, tensor perturbations and the Boltzmann equations for photons, finalizing with a resume of the results at perturbative level. Finally, in Sec. \ref{CO} we give some conclusions and outlooks.

\section{Review on Background Cosmology with UG} \label{Sec:Review}

Unimodular gravity can be described by the following field equation
\begin{equation}
R_{\mu\nu}-\frac{1}{4}g_{\mu\nu}R=8\pi G\left(T_{\mu\nu}-\frac{1}{4}g_{\mu\nu}T\right), \label{UGfield}
\end{equation}
where all the tensors are the standards of GR and $G$ is the Newton's gravitational constant. This equation can be deduced formally by Einstein-Hilbert action under the consideration $\sqrt{-g}=\xi$, where $\xi$ is a constant.

In order to study the background cosmology, we consider an isotropic, homogeneous and flat Friedmann-Lemaitre-Robertson-Walker (FLRW) metric\footnote{An important clarification is that a change of variables  (such as $a\to b^{1/4}$ and $dt\to b^{-3/4}d\tau$), applied by  \cite{Alvarez:2015sba} to FLRW metric to maintain the UG constraint, does not modify the mathematical consistency and it only impacts the physical interpretation (see details in \cite{Garcia-Aspeitia:2019yni} appendix).}, $ds^2=-dt^2+a(t)^2d\vec{x}^2$, where $a$ is the scale factor, the perfect fluid energy momentum tensor is written as $T_{\mu\nu}=pg_{\mu\nu}+(\rho+p)u_{\mu}u_{\nu}$, where $p$, $\rho$ and $u_{\mu}$ are the pressure, density and four-velocity of the fluid respectively. Hence, the integrability for FLRW metric in UG give us  \cite{Ellis,Gao:2014nia}
\begin{equation}
\dot{H}=\frac{\ddot{a}}{a}-H^2=-4\pi G\sum_i(\rho_i+p_i), \label{Friedmann}
\end{equation}
where the dots stands for time derivative.
In addition, a general conservation for UG theory is now written in the form
\begin{equation}
\nabla^{\mu}[32\pi GT_{\mu\nu}-(R+8\pi GT)g_{\mu\nu}]=0. \label{Eq:noncons}
\end{equation}
Without independently assuming the energy momentum conservation ($\nabla^{\mu}T_{\mu\nu}=0$), the Eq. \eqref{Eq:noncons} introduces new Friedmann, acceleration and fluid equations coupled with third order derivatives in the scale factor due to second order derivatives in the Ricci scalar.
Hence, in the case of non traditional conservation of the energy-momentum tensor, Eq. \eqref{Eq:noncons} must be solved to obtain the characteristic fluid equation. Hence, solving for \eqref{Eq:noncons} under a FLRW metric and perfect fluid we have
\begin{equation}
\sum_i\left[\frac{d}{dt}(\rho_i+p_i)+3H(\rho_i+p_i)\right]=\frac{H^3}{4\pi G}(1-j), \label{chida2}
\end{equation}
where the sum is over all the species in the Universe and $j\equiv\dddot{a}/aH^3$ is the Jerk parameter (JP) \cite{Zhang:2016,Mamon:2018dxf}, well known in cosmography and proposed by \cite{Garcia-Aspeitia:2019yni} for the study of UG. The jerk parameter is the key ingredient to avoid problems with the initial-values related to third order derivatives, discarding spurious solutions with no physical meaning, because of its well-known behavior. The idea of the jerk parameter will be discussed later.

On the other hand, the integral-transcendent-Friedmann equation can be computed with the help of Eq. \eqref{Friedmann} and \eqref{chida2}, obtaining the Friedmann equation for UG as 
\begin{equation}
H^2=\frac{8\pi G}{3}\sum_i\rho_i+H^2_{corr}. \label{Frie}
\end{equation}
In addition, the acceleration equation is deduced from \eqref{Friedmann}, obtaining
\begin{equation}
\left(\frac{\ddot{a}}{a}\right)=-\frac{4\pi G}{3}\sum_i\left(\rho_i+3p_i\right)+H_{corr}^2, \label{acc}
\end{equation}
where the non-canonical extra term in Eqs. \eqref{Frie} and \eqref{acc}, i.e. the UG correction to the Friedmann and acceleration equations, is defined in the form
\begin{equation}
H_{corr}^2\equiv\frac{8\pi G}{3}\sum_ip_i+\frac{2}{3}\int_{a_{ini}}^{a(t)} H(a^{\prime})^2[j(a^{\prime})-1]\frac{da^{\prime}}{a^{\prime}}, \label{HUG}
\end{equation}
where the sum runs over the different species in the Universe and $a_{ini}$ is some constant initial value, which will be used as free parameter as discussed later. 
The constant term is possible because in UG we have $\sqrt{-g}=\xi$, being $\xi$ a constant (see \cite{Ellis}), and thus allowing $a_{ini}$ to be a free parameter. The integral start from late epochs (today) and explore any value in all epoch of the Universe evolution through $a(t)$.

Although the Universe acceleration happens when $\int H^2(j-1)a^{-1}da>2\pi G(\rho+p)$, which allows common fluids to accelerate the Universe, the presence of the scale factor in the expression restrict the acceleration epoch.

According to \cite{Garcia-Aspeitia:2019yni}, it is plausible to consider an ansatz for the JP in terms of the redshift with the following characteristics
\begin{equation}
j(z)=\frac{9(1+w)w}{2E(z)^2}\Omega_{0i}(z+1)^{3(w+1)}+1, \label{JX}
\end{equation}
where $w$ is the EoS for any fluid. This expression for the jerk is not unique (other equations could be considered) and it is proposed  to fulfill Eq. \eqref{chida2}. Notice how the function reproduce both the acceleration (driven by a cosmological constant) and the matter stages. If we additionally choose $\Omega_{0i}\to\Omega_{0r}$ and $w\to w_r=1/3$, this expression for the jerk will also reproduce the radiation stage and, in combination with the other stages, the $\Lambda$CDM jerk in all eras \cite{Garcia-Aspeitia:2019yni}. Physically, the jerk parameter is a guide to reproduce the different stages because it involves third order derivatives of the scale factor. It also produce the expected expansion for the enthalpy ($\propto a^{-4}$) as suggested in \cite{Velten:2021xxw}. In summary, this choice for the jerk parameter not only recreates the benefits of $\Lambda$CDM model and alleviate the problems associated with the initial conditions, but also unveils some aspects of the nature of CC-like, i.e. introducing advantages over the standard cosmological model.

As mentioned previously, a crucial characteristic is that this jerk parameter could come from third order derivatives in the equations of a more fundamental theory of space-time (allowing the violation of the energy-momentum conservation) or by effects in the measurement problem in quantum mechanics (QM) as argued by \cite{Perez:2018wlo}. For example,  regarding space-time, the jerk could be originated by the causal sets (CS) approach to quantum gravity \cite{Josset:2016}, where in the cosmological context the diffusion in the phase-space leads to the violation of the energy-momentum conservation. Alternatively, the jerk parameter could emerge from the continuous spontaneous localization (CSL), also related to the measurement problem in QM, which is connected to energy divergences in quantum collapse models, establishing a correlation with UG as it is shown in \cite{Josset:2016,Ballentine,Bassi_2005}. In addition, Corral et al. \cite{Corral:2020} tackle the problem by arguing that the jerk parameter presented in Eq. \eqref{JX} could be a diffusion parameter $Q$ in the equation of state, which would mimic the Universe acceleration.

From the previous equation and Eq. \eqref{Frie} it is possible to deduce
\begin{eqnarray}
E(z)^2&=&\Omega_{0m}(z+1)^3+\Omega_{0r}(z+1)^4\nonumber\\&&+\Omega_{0exs}(z_{ini}+1)^4, \label{HX} 
\end{eqnarray}
where $E(z)\equiv H(z)/H_0$, $\Omega_{0exs}\equiv w_r\Omega_{0r}$ and $z_{ini}$ comes from the constant term  in the Eq. \eqref{HUG} integral. It is worth mentioning that we have the following relation 
\begin{equation}\label{eq:flatness}
    z_{ini} = -1 + \left( \frac{1-\Omega_{0m}-\Omega_{0r}}{w_r\Omega_{0r}}\right)^{1/4},
\end{equation}
coming from the flatness condition. Notice that the source of the Universe acceleration is the constant term in the previous equation (Eq. 9), which is a combination of $\Omega_{0exs}$ and $z_{ini}$, in where we can naturally relate $\Omega_{0\Lambda}\to\Omega_{0exs}(z_{ini}+1)^4$. Since our choice in Eq. \eqref{JX} depends on the EoS and the energy density parameter of the radiation, the constant inherits those terms. An upper limit for $z_{ini}$ can be established through the equation
\begin{equation}
    z_{ini}+1<\Omega_{0exs}^{-1/4}, \label{constr}
\end{equation}
hence, for $\Omega_{0exs}=8.23\times10^{-6}h^{-2}(1+0.2271g_*)$, being $g_*=3.04$ the standard number of relativistic species \cite{Komatsu:2011}, and $h$ is the dimensionless Hubble constant. This result is dictated under the assumption that $\Omega_{0\Lambda}=\Omega_{0exs}(z_{ini}+1)^4<1$, obtaining an upper limit for $z_{ini}$ to be $z_{ini}<12.54$ at $95\%$  confidence level (CL), using Planck mission value $h=0.6766^{ + 0.0042}_{- 0.0042}$  \cite{Planck:2018} (notice that the result for this constriction is strongly dependent of what $h$ value is considered).

$\Lambda$CDM model (which our theory mimics) considers the reionization era in the region $6<z<20$, predicting a ionization fraction of $\sim0.2$ at $z\sim12$ (see Ref. \cite{Paoletti:2020ndu} for details). On the other hand, in order to agree with the expected value for $\Omega_{0\Lambda}$ in GR, we require that $z_{ini}^{CMB}=11.15^{+0.01}_{-0.02}$. From here, we conclude that the CC-like origin is approximately at the epoch of reionization in the late Universe\footnote{In Ref. \cite{Perez:2018wlo}, the CC can be traced from electroweak epoch, due to some specific quantum violation related to the energy-momentum tensor.}, which is one of the most important results of this theory.
It is worth to notice that these results are under the assumption that $\Omega_{0\Lambda}=0.6889^{+ 0.0056}_{- 0.0056}$ and $h=0.6766^{+0.0042}_{-0.0042}$ are taken from Ref. \cite{Planck:2018}. Having said that, if we use the value of Hubble constant from SnIa data measured by \cite{Riess:2016jrr}  ($h=0.7422^{+0.0182}_{-0.0182}$), the result differs from the one from Planck as expected. Hence, using this $h$ estimation and assuming that $\Omega_{0\Lambda}$ coincide with the Planck value\footnote{ In general, this assumption is not true, however Riess et al. \cite{Riess:2019cxk} does not report the expected value for $\Omega_{0\Lambda}$ in the Supernovaes observations.}, we obtain that $z_{ini}^{SnIa}$ moves to $11.43^{+0.03}_{-0.06}$ ($1\sigma$ CL), not too far from the  previous value of $z_{ini}^{CMB}=11.15$. 

\section{Background Cosmological constrictions} \label{cosmological}

At background level the parameters $z_{ini}$ and $h$ can be constrained by performing a Bayesian Markov Chain Monte Carlo (MCMC) analysis employing OHD, SnIa, CMB and BAO data.

To perform the MCMC analysis, we use the \textit{emcee} Python module \cite{Foreman:2013}  choosing a flat prior over all the parameters in the range $h:[0.2,1.0]$, $\Omega_b h^2:[0, 0.04]$, $\Omega_{0m}:[0,1]$. To establish a bound over the parameter $z_{ini}$, we use Eq. \eqref{eq:flatness}. We set a burn-in phase to achieve the convergence according to Gelman-Rubin criteria \cite{Gelman:1992} and $4000$ MCMC steps with $250$ walkers.

Then, we build a Gaussian log-likelihood as the merit-of-function to minimize $
-2\log(\mathcal{L}_{\rm data})\varpropto \chi^2_{\rm data}$, for each dataset mentioned previously. Additionally, a joint analysis can be constructed through the sum of them, i.e.,
\begin{equation}
    \chi^2_{\rm Joint}=\chi^2_{\rm SnIa}+\chi^2_{\rm OHD}+\chi^2_{\rm CMB}+\chi^2_{\rm BAO},
\end{equation}
where subscripts indicate the observational measurements under consideration. The rest of the section is devoted to describe the different cosmological observations.

\subsection{ Supernovae Type Ia}

The largest compilation provided by Ref. \cite{Scolnic:2017caz}, contains the observations of the luminosity modulus from 1048 SnIa spanned in the redshift region $0.01<z<2.3$. The $\chi^2$-function is constructed as \cite{Conley_2010}
\begin{equation}
    \chi_{Pan_{\rm SnIa}}^{2}=a +\log \left( \frac{e}{2\pi} \right)-\frac{b^{2}}{e}, \label{fPan}
\end{equation}
where $a=\Delta\boldsymbol{\tilde{\mu}}^{T}\cdot\mathbf{C_{P}^{-1}}\cdot\Delta\boldsymbol{\tilde{\mu}},\, b=\Delta\boldsymbol{\tilde{\mu}}^{T}\cdot\mathbf{C_{P}^{-1}}\cdot\Delta\mathbf{1}$,\, $e=\Delta\mathbf{1}^{T}\cdot\mathbf{C_{P}^{-1}}\cdot\Delta\mathbf{1}$, $\Delta\boldsymbol{\tilde{\mu}}=\tilde{\mu}_{\mathrm{th}}-\tilde{\mu}_{\mathrm{obs}}$ is the vector of residuals between the model distance modulus and the observed one $\tilde{\mu}_{\mathrm{obs}}$, $\Delta\mathbf{1}$ is the unit vector, and $\mathbf{C_{P}^{-1}}$ is the inverse of the covariance matrix. The theoretical distance modulus is estimated by
\begin{equation}
    \tilde{\mu}_{\mathrm{th}}(z)=\mathcal{M}+5\log_{10}[d_L(z)/10\, pc],
\end{equation}
where $\mathcal{M}$ is a nuisance parameter that has been marginalized in \eqref{fPan}, and $d_L(z)$ is the dimensionless luminosity distance given by the following equation 
\begin{equation}\label{eq:dL}
    d_L(z)=(1+z)c\int_0^z\frac{dz^{\prime}}{H(z^{\prime})},
\end{equation}
where $c$ is the light velocity.

\subsection{Observational Hubble Data}

Another important sample, is the Observational Hubble Data (OHD) which consist of cosmological model independent measurements of the Hubble parameter $H(z)$. We consider the OHD compilation provided by \cite{Magana:2017nfs} comprised by 51 points given by the differential age (DA) tool (20 points) and BAO measurements (31 points) within the redshift region $0<z<2.36$. The $\chi^2$-function for OHD can be written as
\begin{equation}
    \chi^2_{{\rm OHD}}=\sum_i^{51}\left(\frac{H_{th}(z_i)-H_{obs}(z_i)}{\sigma^i_{obs}}\right)^2,
\end{equation}
where $H_{th}(z)$ and $H_{obs}(z_i)$ are the theoretical and observational Hubble parameter at the redshift $z_i$, and $\sigma_{obs}^i$ is the observational error.

\subsection{CMB from Planck 2018 measurements}
Recently, \cite{Basak:2015} performed a further perturbative analysis of the UG model. They found that, at first and second order of perturbative level, there are no differences between $\Lambda$CDM and UG, hence the UG theory is not excluded by  measuring either the growth factor, CMB lensing or the integrated Sachs-Wolfe effect. Therefore, as a first approach, we consider the CMB posterior distances from $\Lambda$CDM to test this model. 

We use the shift parameter, $R=1.7502\pm0.0046$, the acoustic scale, $l_{A}=301.471^{+0.089}_{-0.090}$, 
and $\Omega_{b0}h^{2}=0.02236\pm0.0015$ obtained for a flat $\Lambda$-cold dark matter model \cite{Chen:2018dbv}. Thus, the figure-of-merit is built as

\begin{equation}
    \chi^2_{\rm CMB}=V_{\rm CMB}\cdot \rm{Cov}_{\rm CMB}^{-1}\cdot V_{\rm CMB}^T,
\end{equation}
where $V_{\rm CMB}$ is

\begin{equation}
 V_{\rm CMB} =\left(
 \begin{array}{c}
 R^{th} - 1.7502\\
 l_A^{th} - 301.147 \\
 \Omega_{b}h^{2 th} - 0.02236
\end{array}\right),
\end{equation}
the superscripts $th$ refer to the theoretical values, 
and $\rm Cov_{CMB}^{-1}$ represents the inverse of
\begin{equation}
\rm{Cov_{CMB}} = 10^{-8}\left(
\begin{array}{ccc}
2116.00 & 18938.20 & -45.54   \\
18938.20 & 801025.00 & -443.03\\
-45.54 & -443.03 & 2.25
\end{array}\right).
\label{eq:invcovcmb}
\end{equation}
which is the  covariance matrix for $ V_{\rm CMB}$.

\subsection{Baryon Acoustic Oscillations}

Baryon Acoustic Oscillations (BAO) are considered as standard rulers, being primordial signatures of the interactions between baryons and photons in a hot plasma on the matter power spectrum in the pre-recombination epoch. In Ref. \cite{nunes:2020}, the authors collected 15 transversal BAO scale measurements, obtained from luminous red galaxies located in the region $0.110<z<2.225$.

In order to be used as a method to constrain cosmological models, it is useful to build the $\chi^2$-function as
\begin{equation}
\chi^2_{\rm BAO} = \sum_{i=1}^{15} \left( \frac{\theta_{\rm BAO}^i - \theta_{th}(z_i) }{\sigma_{\theta_{\rm BAO}^i}}\right)^2\,,
\end{equation}
where $\theta_{\rm BAO}^i$ is the BAO angular scale and its uncertainty $\sigma_{\theta_{\rm BAO}^i}$ measured at $z_i$. The theoretical counterpart, $\theta_{th}$, is estimated through the following equation
\begin{equation}
    \theta_{th}(z) = \frac{r_{drag}}{(1+z)D_A(z)}\,.
\end{equation}
In the latter, $r_{drag}$ is defined by the sound horizon at baryon drag epoch and $D_A=d_L(z)/(1+z)^2$ is the angular diameter distance at $z$, with $d_L(z)$ defined in equation \eqref{eq:dL}. In addition, we use the $r_{drag}=147.21 \pm 0.23$ reported by \cite{Planck:2018}.

\begin{table}
\caption{Mean values for the 
model parameters ($h$, $z_{ini}$) and $\chi^2_{min}$, derived from each data set and the joint analysis.}
\resizebox{0.4\textwidth}{!}{
\begin{tabular}{|cccc|}
\multicolumn{4}{c}{}\\
\hline
Data set & $\chi^{2}_{min}$& $h$ & $z_{ini}$ \\
\hline
OHD  &   $22.0$ & $0.709^{+0.016}_{-0.016}$  & $11.788^{+0.237}_{-0.250}$ \\
SnIa & $1036.0$ & $0.602^{+0.270}_{-0.272}$  & $10.623^{+2.366}_{-3.021}$ \\
CMB  & $0.0001$ & $0.678^{+0.006}_{-0.006}$ & $11.259^{+0.091}_{-0.092}$  \\
BAO  & $12.9$ & $0.701^{+0.031}_{-0.033}$ &  $10.847^{+0.979}_{-1.383}$  \\
Joint & $1097.6$ & $0.692^{+0.005}_{-0.005}$ & $11.473^{+0.074}_{-0.073}$ \\
\hline
\end{tabular}}
\label{tab:par}
\end{table}

\begin{figure}
\centering
\par\smallskip
{\includegraphics[width=0.35\textwidth]{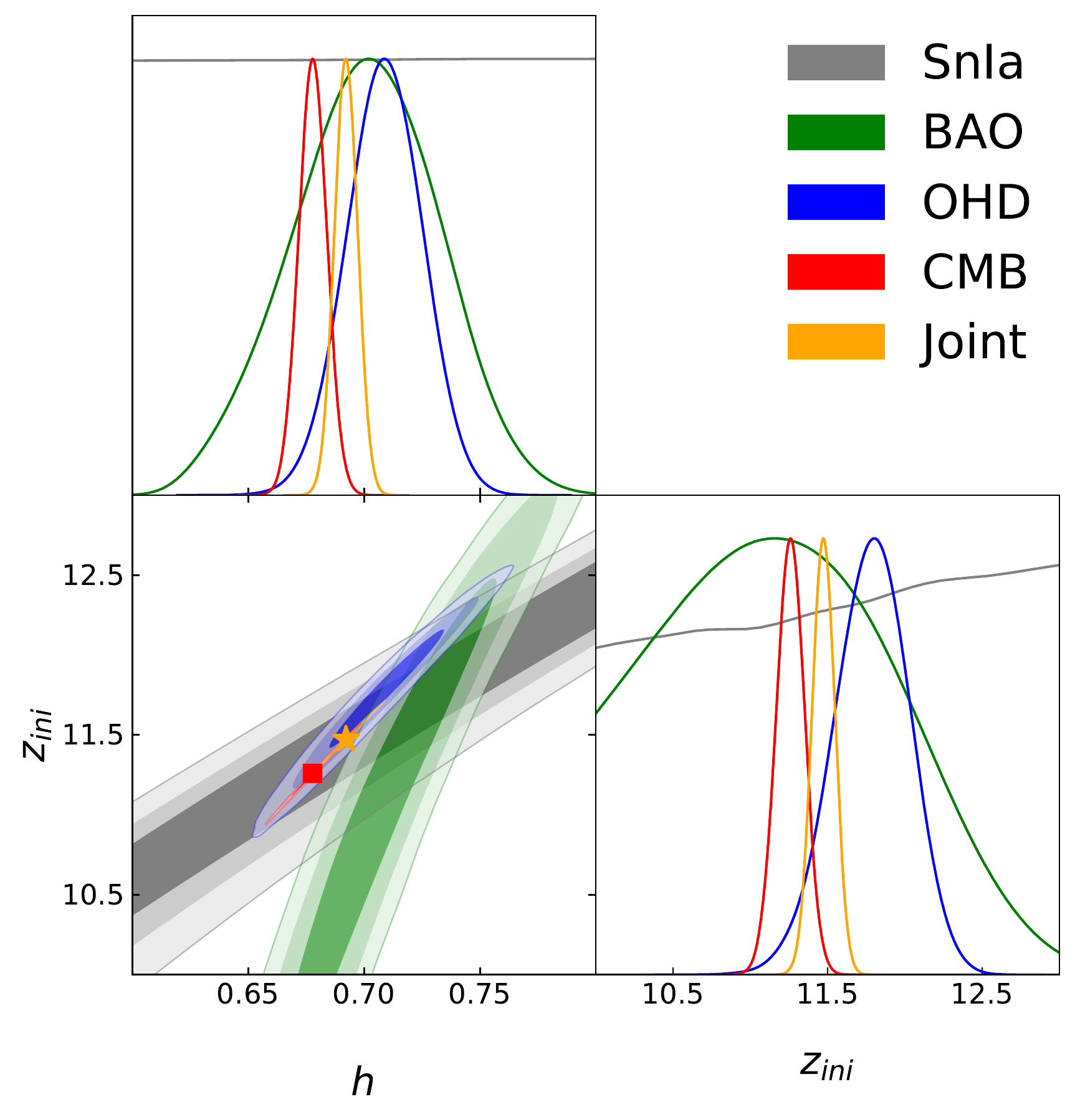}}
\caption{1D marginalized posterior distributions and the 2D $68\%$, $95\%$, $99.7\%$ of CL for the $h$, and $z_{ini}$ parameters of the UG model. The star (square) marker represents the best fit value of Joint (CMB) data}
\label{Figure:CosmologicalData}
\end{figure}

\subsection{Results at Background Level}

Table \ref{tab:par} provides the mean values for $h$ and $z_{ini}$ obtained from each data set and a joint analysis. Furthermore, Figure \ref{Figure:CosmologicalData} presents the 1D marginalized posterior distributions and 2D at $68\%$, $95\%$, $99.7\%$ of CL for the UG parameters. The results for the five cases confirm that $z_{ini}\sim12$; hence, we conclude that the CC-like should emerge in the reionization epoch (see \cite{Paoletti:2020ndu}) in contrast to the standard cosmological model where the CC is always present.

More precisely, the joint analysis estimates $z_{ini}=11.47^{+0.07}_{-0.07}$, while $z_{ini} = 11.15^{+0.01}_{-0.02}$ and  $11.43^{+0.03}_{-0.06}$ are those obtained with the $h$ values from Planck, and SnIa measurements respectively. Note that the $4.6\sigma$ between CMB and the joint analysis is due that we choose the value of $h$ given by Riess et al. \cite{Riess:2019cxk} instead the presented by Planck \cite{Planck:2018}. As mentioned before, these values lie on the reionization region ($6<z<20$) \cite{Bouwens:2015vha} (see \cite{Paoletti:2020ndu} for other regions with diverse conditions), and are also in agreement with recent measurements \cite{Liu:2019awk} by the Shaped Antenna of the background Radiation Spectrum (SARAS), which ruled out reionization in the region $10 < z < 6$. 
 Several important events happened at that time such as the apparition of high-redshift galaxies \cite{Bouwens:2015vha} or the birth of the first stars in the Universe, the so-called Population III stars \cite{Magg,Glover,Trenti}. In that epoch, the combination of the above mentioned mechanisms reionize the intergalactic medium initiating the process of metallicity in the Universe (and bringing the end of the so-called dark ages epoch). We remark that this epoch is particularly important because non linear physics processes are involved. We speculate that the friction of non linear structures (galaxies, clusters, etc) with the granularity of the space-time could have helped to the emergence of small violation of the energy-momentum conservation which can be accounted as the energy density of the CC. For example, the collapsed structures in the epoch of reionization are a way to  brake the symmetry of the space-time, which is extensively explored by authors like in Ref. \cite{Lombriser:2018aru}. Furthermore, CSL or CS, processes could be other reasons to the apparition of the CC, as we discussed earlier.

\section{Equations associated to linear perturbations of UG} \label{PERT}

We dedicate this section to elucidate the equations associated to the linear perturbations for UG, starting with the field equations and later on with the Boltzmann equations, centering our attention in the photons case because, as discussed in the previous section, this is the equation coupled with the jerk parameter.

In order to study the equations associated to the linear perturbations, we rewrite Eq. \eqref{UGfield} in the form
\begin{equation}
    G_{\mu\nu}+\frac{1}{4}(R+8\pi GT)g_{\mu\nu}=8\pi GT_{\mu\nu}, \label{UGRewriten}
\end{equation}
where $G_{\mu\nu}\equiv R_{\mu\nu}-\frac{1}{2}g_{\mu\nu}R$, is the Einstein tensor.

We start using the conformal Newtonian gauge line element given by the components
\begin{eqnarray}
&&g_{00}(t,{\bf x})=-1+h_{00}(t,{\bf x}), \\
&&g_{0i}(t,{\bf x})=a(t)h_{0i}(t,{\bf x}), \\
&&g_{ij}(t,{\bf x})=a^2(t)[\delta_{ij}+h_{ij}(t,{\bf x})],
\end{eqnarray}
where $h_{\mu\nu}\ll1$, are metric perturbations.

\subsection{Scalar Perturbations} \label{SP}

For this particular case, we reduce the above equations to $h_{00}(t,{\bf x})=-2\Psi(t,{\bf x})$, $h_{0i}=0$ and $h_{ij}(t,{\bf x})=2\delta_{ij}\Phi(t,{\bf x})$, where $\Psi(t,{\bf x})$ and $\Phi(t,{\bf x})$ are the Newtonian and curvature potential, respectively.

The first equation that we present is associated with the $00$ component, therefore we have the following parts

\begin{eqnarray}
    &&\delta G^0_0=-6H\dot{\Phi}+6 H^2\Psi-\frac{2k^2}{a^2}\Phi, \label{G00}\\
    &&\delta R=-12\Psi\left(H^2+\frac{\ddot{a}}{a}\right)+\frac{2k^2}{a^2}\Psi+6\ddot{\Phi}-6H(\dot{\Psi}-4\dot{\Phi})\nonumber\\&&+\frac{4k^2}{a^2}\Phi,\label{R}\\
    &&\delta T^0_0=-\sum_i\rho_i\delta_i-\rho_{\gamma}\Theta_0-\rho_{\nu}\mathcal{N}_0,\label{T00}\\
    &&\delta T=-3\sum_i\rho_i\delta_i-2(\rho_{\gamma}\Theta_0+\rho_{\nu}\mathcal{N}_0), \label{TEsc}
\end{eqnarray}
where $k$ is the wave-number related to the perturbation scale, $\partial/\partial x^j\to ik_j$, $\delta_i\equiv\delta\rho_i/\rho_i$ are the fractional overdensities for DM and baryons, while $\Theta_0$ and $\mathcal{N}_0$ are the monopolar contributions for photons and neutrinos perturbations. Notice also that $\delta$ indicates that we are working with first order terms (linear perturbations).

Combining previous equations in the $00$ component of \eqref{UGRewriten} at perturbative level we have

\begin{eqnarray}
    &&-3\mathcal{H}\Phi'+3 \mathcal{H}^2\Psi-k^2\Phi+\frac{1}{2}\Lambda(\eta,k)=-4\pi Ga^2\Big[\sum_i\rho_i\delta_i+\nonumber\\&&\rho_{\gamma}\Theta_0+\rho_{\nu}\mathcal{N}_0\Big]
\end{eqnarray}
where 

\begin{eqnarray}
    &&\Lambda(\eta,k)=3\Psi\left(\mathcal{H}^2+\frac{a''}{a}\right)-\frac{k^2}{2}\Psi-\frac{3}{2}\Phi''+\frac{3}{2}\mathcal{H}(\Psi'-4\Phi')\nonumber\\&&-k^2\Phi+2\pi Ga^2\left[3\sum_i\rho_i\delta_i+2(\rho_{\gamma}\Theta_0+\rho_{\nu}\mathcal{N}_0)\right], \label{LambdaPert}
\end{eqnarray}
here the equations are transformed to the conformal time through the relation $dt=ad\eta$, primes denote derivatives with respect to $\eta$ ($'=\partial_{\eta}$) and $\mathcal{H}$ is the Hubble parameter in conformal time. Notice that when $\Lambda(\eta,k)=\mathcal{C}$, where $\mathcal{C}\ll1$ is a constant small perturbation, the equation that rise from GR plus a constant should be recovered. However, this is no trivial, because the previous imposition, generates another constraint over the functions that contains \eqref{LambdaPert}. The hypothesis and mathematical arguments used to infer $\Lambda(\eta,k)=\mathcal{C}$ needs the Boltzmann equations discussed later.

On the other hand, for the $ij$ equations we use the longitudinal traceless (LT) part of $G^i_j$ denoted by $\hat{k}_i\hat{k}^j-(1/3)\delta^j_i$. Considering that the term $(R+8\pi GT)/4$ contains a product with $\delta^i_j$, this is neglected by the LT consideration, therefore the final form is equal to those expected by GR 
\begin{equation}
    k^2(\Phi+\Psi)=-32\pi Ga^2[\Theta_2\rho_{\gamma}+\mathcal{N}_2\rho_{\nu}], \label{LT}
\end{equation}
where $\Theta_2$ and $\mathcal{N}_2$ are the quadrupole contributions of photons and neutrinos, respectively.

\subsection{Tensor Perturbations} \label{TP}

For tensor perturbations, we begin describing the metric components as $h_{00}=-1$, $h_{0i}=0$ and

\begin{equation}
    \delta g_{ij}(t,{\bf x})=a(t)^2h_{ij}(t,{\bf x}),
\end{equation}
where
\[ h_{ij}=\left( \begin{array}{ccc}
h_+ & h_{\times} & 0 \\
h_{\times} & -h_+ & 0 \\
0 & 0 & 0 \end{array} \right),\]
being $h_{+},h_{\times}\ll1$ two components of the divergenceless-traceless symmetric tensor chosen in the $x-y$ plane. The subscripts $+$ and $\times$ indicates the wave polarizations. Therefore, the components to construct Eq. \eqref{UGRewriten} are
\begin{eqnarray}
    &&\delta G^i_j=\delta R^i_j=\delta^{ik}\left[\frac{1}{2}h_{kj,00}+\frac{k^2}{2a}h_{kj}+\frac{3}{2}Hh_{kj,0}\right],\\
    &&\delta R=0,
\end{eqnarray}
the $ij$ components of the energy momentum tensor is as always (avoiding anisotropic stress contributions)
\begin{equation}
    T^i_j(t,{\bf x})=\sum_sg_s\int\frac{d^3p}{(2\pi)^3}\frac{p^ip_j}{E_s(p)}f_s({\bf x},{\bf p},t),
\end{equation}
where $f_s({\bf x},{\bf p},t)$ is the distribution function, $g_s$ values are the degrees of freedom of the fluids, $p^i$ and $E(p)$ are the three momenta and energy of the particles involved, respectively. Moreover, $\delta T$ is the same as that shown in Eq. \eqref{TEsc}. Following the traditional guidelines we have that

\begin{equation}
    a^2[\delta G^1_1-\delta G^2_2]=h_i''+2\mathcal{H}h'_i+k^2h_i,
\end{equation}
where $i=+,\times$, and 

\begin{equation}
    8\pi G\left[\delta T^1_1-\delta T^2_2-\frac{1}{4}T(g^1_1-g^2_2)\right]=0,
\end{equation}
thus,
\begin{equation}
   h_i''+2\mathcal{H}h'_i+k^2h_i=0.
\end{equation}
Therefore, the linear tensor perturbations does not show any difference with the one from GR.

\subsection{Boltzmann equations} \label{BE}

As we discussed previously, Eq. \eqref{chida2} is the main  equation for the fluids with the sum indicating that all the components considered in the cosmology should be added. For the matter component (with $p=0$), the well know equation $\dot{\rho}_m+3H\rho_m=0$ is directly recovered, while for the radiation component the jerk is designed {\it ad-hoc} in such a way that the equation $\dot{\rho}_r+4H\rho_r=0$ for relativistic particles is recovered. Thus, it is expected that linear perturbations of Boltzmann equation for photons should carry the information encrypted in the $j$ term, even in the free streaming epoch when Compton process could be safely neglected. In this epoch, the particle density is low, implying that the optical depth is negligible ($\tau\ll1$) and defined as $\tau=\int n_e\sigma_Tad\eta$, being $n_e$ the electron number density and $\sigma_T$ the Thompson cross section and hence, the interaction part described by the collision function $C[f({\bf p})]$ will not play a role.

From Eq. \eqref{Eq:noncons} it is possible to observe that $\nabla^{\mu}T_{\mu\nu}=\partial_{\nu}\mathcal{F}(R,T)$, being $\partial_{\nu}\mathcal{F}(R,T)$ a function that involves derivatives of the Ricci and energy-momentum scalars (the function absorbs the factor $(32\pi G)^{-1}$ ), which implies higher order derivatives as it is shown in equation \eqref{chida2} encrypted in the jerk parameter. However, notice that the derivatives are only depending on time because $\rho$ and $p$ only depends on $t$, therefore we have $\partial_t\mathcal{F}(R,T)$ and then we arrive to Eq. \eqref{chida2}.

Consequently, if photons carry the violations to the energy momentum tensor\footnote{At background level, we only claim that radiation (relativistic particles) carry the energy-momentum violation, but we do not specify if those are the photons or the neutrinos. Here, we are adding another extra hypothesis assuming that photons specifically carry energy-momentum violations.}, we expect that the linear perturbed equations in conformal time with $\tau\ll1$, will take the form
\begin{equation}
    \Theta'+ik\mu\Theta+\Phi'+ik\mu\Psi=\partial_{\eta}\tilde{\mathcal{F}}(\delta R,\delta T), \label{photonEq}
\end{equation}
where $k\mu={\bf k}\cdot\hat{{\bf p}}$, and $\hat{\bf p}$ is the momentum direction, $\Theta\equiv\delta T/T$ is the photon perturbations, which cause differences in the temperature, and $\tilde{\mathcal{F}}$ is the perturbed part of the function $\mathcal{F}$, involving $\delta R$ and $\delta T$ given by Eqs. \eqref{R} and \eqref{TEsc} for the scalar perturbations. Then, we have 
for a scalar energy-momentum tensor composed only by photons

\begin{eqnarray}
    &&\partial_{\eta}\tilde{\mathcal{F}}(\delta R,\delta T)=(32\pi G)^{-1}\partial_{\eta}\Big[3\Psi\left(\mathcal{H}^2+\frac{a''}{a}\right)-\frac{k^2}{2}\Psi\nonumber\\&&-\frac{3}{2}\Phi''+\frac{3}{2}\mathcal{H}(\Psi'-4\Phi')-k^2\Phi+4\pi Ga^2\rho_{\gamma}\Theta_0\Big]\nonumber\\&&=(32\pi G)^{-1}\partial_{\eta}\Lambda(\eta,k). \label{DerFTilde}
\end{eqnarray}
Notice that the derivatives involve the conformal time parameter, then, from Eq. \eqref{R} we will have higher order derivatives for $a$ and $\Phi$. Thus, if we want to obtain the standard perturbations for GR, the equation \eqref{JX} is not useful anymore. In the same way, we need to propose a new jerk parameter in order to face the problems associated with third order derivatives encoded in the Boltzmann equation for photons.

As a final remark, we expect that the Boltzmann equations for the remaining species must be the same as these obtained in the standard cosmology. Thus, the mathematical structure for the UG equations suggest that only radiation (at background level) and photons (at perturbative level) are the only species coupled with the non conservative term. While we increase the energy, the Compton scattering appears and therefore, Eq. \eqref{photonEq} must contain the term $-\tau'[\Theta_0-\Theta+u_b\mu]$, where $u_b$ is the bulk velocity. Regarding the initial conditions associated to the inflationary process, those must be studied in a future work because, unlike the GR case, the analysis requires extra contributions.

\subsection{Results at Perturbative Level}

UG equations at perturbative level are presented, implementing the conformal Newtonian gauge line element, showing the scalar and tensor perturbations as well as the Boltzmann equation for photons. It is notorious that the scalar perturbations for $00$ contains a term $\Lambda(\eta, k)$, not expected in standard GR, promoting another constriction into UG equations. Regarding the $ij$ component, for scalar perturbations and due to the LT consideration, Eq. \eqref{LT} maintains the same form as the one shown in GR. In the case of linear tensor perturbations, we conclude that the equations do not change, having the same result as predicted by GR.

On the other hand, it is possible to deduce the Boltzmann equation for the photons, assuming a negligible optical depth, i.e. no Compton scattering. The left side of the equation is the same as the one expected in standard cosmology, meanwhile the right side will contain a function that depends on $\delta R$, $\delta T$ and higher order conformal temporal derivatives that now also involves the spatial curvature $\Phi$. In this case, Eq. \eqref{JX} is not useful as a guide to avoid problems with high order derivatives. The classical way to solve this problem is to assume Eq \eqref{DerFTilde} is equal to zero, therefore $\Lambda(\eta,k)=\mathcal{C}$, which justify the previously mentioned assumption.

Therefore, UG provides an extra equation which reads

\begin{eqnarray}
    &&3\left(\mathcal{H}^2+\frac{a''}{a}-\frac{k^2}{6}\right)\Psi-\frac{3}{2}\Phi''+\frac{3}{2}\mathcal{H}(\Psi'-4\Phi')\nonumber\\&&-k^2\Phi+4\pi Ga^2\rho_{\gamma}\Theta_0=\mathcal{C}. \label{ExtraEq}
\end{eqnarray}
We will call it the {\it constriction equation} for UG. 

Conversely, we can face Eq. \eqref{DerFTilde} which implies dealing with higher order derivatives. In this case, the strategy should be the same  as the one for the homogeneous Universe, i.e. find a well known function as a guide (the jerk function). A direct manipulation of Eq. \eqref{photonEq} generates the following integro-differential equation, as it happens for the background case, in the following way 

\begin{eqnarray}
    &&ik\mu\int_{\eta_0}^{\eta}(\Theta+\Psi)d\eta=(32\pi G)^{-1}\Big[3\Psi\left(\mathcal{H}^2+\frac{a''}{a}\right)\nonumber\\&&-32\pi G(\Theta+\Phi)-\frac{k^2}{2}\Psi-\frac{3}{2}\Phi''+\frac{3}{2}\mathcal{H}(\Psi'-4\Phi')\nonumber\\&&-k^2\Phi+4\pi Ga^2\rho_{\gamma}\Theta_0\Big], \label{InDiffPert}
\end{eqnarray}
which must be solved in combination with the fluids and Einstein equations. 

\section{Conclusions and Outlooks} \label{CO}

Unimodular gravity, and its particular and natural characteristic of violating the traditional form of the energy-momentum conservation, not only suggests the apparition of the CC-like term together with its radiation inherited characteristics (which is assumed coupled with the jerk parameter), but also the redshift at which it emerges. In this scenario, the CC-like arises during the epoch of reionization, when the newly form galaxies and stars of the Population III reionizated the intergalactic medium and, in general, when complex structures played a role in the cosmological dynamics and converting the Universe into a highly non-linear entity.

Towards redshifts of the order $z\gtrsim11$, the CC-like should generate a fingerprint that could be detected in future observations. Hence, a particular form to refute or validate the model is the prediction that CC does not exist in all stages of the evolution of the Universe, but it arises during the epoch of reionization, being the smoking-gun epoch to obtain evidence of CC and elucidate its future implications for the Universe evolution. Models like UG where it is predicted a late CC, are called \textit{emergent} which the capability to alleviate the recent problem related with observations of $H_0$ at late and early cosmological times \cite{Verde:2019ivm}.

We constrained the UG parameters using OHD, SnIa, CMB and BAO data, together with  a joint analysis. Our main result estimates a value for $z_{ini}$ (Joint) of $11.473$, in good agreement with the theoretical values ($11.5$, $11.43$) using $h$ measured by Planck 2018 and \cite{Riess:2019cxk}; strengthening our conclusions that important physics should emerge in the reionization epoch\footnote{Note that the $4.6\sigma$ between CMB and the joint analysis is due that we choose the value of $h$ given by Riess et al. \cite{Riess:2019cxk} instead the presented by Planck \cite{Planck:2018}.}. Furthermore, the estimated values for parameter $z_{ini}$ are in agreement for the data used, which points towards a robust model at the background level and motivates the study of other aspects (such as growth structure at  perturbative level, etc).

Moreover, the Experiment to Detect the Global Epoch of reionization (EDGES, \cite{Bowman:2018yin}) detects an interesting event at a redshift closer to the one we obtain ($z\sim12$): an excess of radiation centered at $z\approx 17$ and spanning a region $20>z>15$. Thus, the EDGES event is consistent with our predictions and it could represent the detection of the CC-like emergence. However, we caution that EDGES result is still in the center of a controversy \cite{Liu:2019awk}, thus, we should wait for new experiments to confirm or refute the observations.

Despite that UG sheds light into the nature of the CC and establishes a redshift for its possible presence, it is not clear why the CC-like emerges at the epoch of reionization. One possibility is that the non linearity in that epoch generates some kind of symmetry breakdown by a still unknown process and this, in turn, leads to the presence of the CC-like originated by the violation of the energy-momentum conservation, resulting in the acceleration of the Universe. As we argue previously, we suspect that non linear physics play an important role in the formation of complex structure (such as stars and galaxies), generating friction with a possible non-continuous space-time and provoking the apparition of the CC-like. The latter assertion could be compatible with CS paradigm or even CSL as we discussed previously.

Cosmological linear perturbations in UG is another subject that is studied in this paper. Under the Newtonian conformal gauge we present scalar and tensor perturbations, together with the Boltzmann equation for photons. Under this scenario, we show how the $00$ component for scalar perturbations, presents an extra contribution call it $\Lambda(\eta,k)$, meanwhile the $ij$ and tensor perturbations remains equal as those obtained by GR. As it is expected, the Boltzmann equation for photons, contains an additional term that contains third order derivatives, not only in the scale factor (as happens in the background) but also to the scalar curvature. The additional hypothesis of consider that Eq. \eqref{DerFTilde} tends to zero, give us the {\it constriction equation} for UG. However, if we face the problem of third order derivatives, it is necessary to solve Eq. \eqref{InDiffPert} (coupled with the other fluids and field equations), which is an integro-differential equation for photons. Formally, it is the Boltzmann equations for photons in UG scenario. As a final remark we mention that it is not possible to generate a numerical analysis of the linear perturbations presented previously and with this, the matter power spectrum, before study the initial conditions in this scenario (inflationary Universe), which will be studied elsewhere.

Finally, the confirmation of this model would not only open the door to understand the current Universe acceleration, but also  could be a piece of evidence for the possible granularity or other primordial effects of the space-time itself.    

\begin{acknowledgements}

The authors acknowledge the enlightening conversations with Luis Ure\~na and Daniel Sudarsky. M.A.G.-A. acknowledges support from SNI-M\'exico, CONACyT research fellow, COZCyT and Instituto Avanzado de Cosmolog\'ia (IAC) collaborations. A.H.A. thanks to the PRODEP project, Mexico for resources and financial support. J.M. acknowledges the support from CONICYT project Basal AFB-170002, V.M. acknowledges the support of Centro de Astrof\'{\i}sica de Valpara\'{\i}so (CAV). J.M., M.A.G.-A. and V.M. acknowledge CONICYT REDES (190147). \\
\end{acknowledgements}

\bibliography{librero1}

\begin{thebibliography}{54}%
\makeatletter
\providecommand \@ifxundefined [1]{%
 \@ifx{#1\undefined}
}%
\providecommand \@ifnum [1]{%
 \ifnum #1\expandafter \@firstoftwo
 \else \expandafter \@secondoftwo
 \fi
}%
\providecommand \@ifx [1]{%
 \ifx #1\expandafter \@firstoftwo
 \else \expandafter \@secondoftwo
 \fi
}%
\providecommand \natexlab [1]{#1}%
\providecommand \enquote  [1]{``#1''}%
\providecommand \bibnamefont  [1]{#1}%
\providecommand \bibfnamefont [1]{#1}%
\providecommand \citenamefont [1]{#1}%
\providecommand \href@noop [0]{\@secondoftwo}%
\providecommand \href [0]{\begingroup \@sanitize@url \@href}%
\providecommand \@href[1]{\@@startlink{#1}\@@href}%
\providecommand \@@href[1]{\endgroup#1\@@endlink}%
\providecommand \@sanitize@url [0]{\catcode `\\12\catcode `\$12\catcode
  `\&12\catcode `\#12\catcode `\^12\catcode `\_12\catcode `\%12\relax}%
\providecommand \@@startlink[1]{}%
\providecommand \@@endlink[0]{}%
\providecommand \url  [0]{\begingroup\@sanitize@url \@url }%
\providecommand \@url [1]{\endgroup\@href {#1}{\urlprefix }}%
\providecommand \urlprefix  [0]{URL }%
\providecommand \Eprint [0]{\href }%
\providecommand \doibase [0]{http://dx.doi.org/}%
\providecommand \selectlanguage [0]{\@gobble}%
\providecommand \bibinfo  [0]{\@secondoftwo}%
\providecommand \bibfield  [0]{\@secondoftwo}%
\providecommand \translation [1]{[#1]}%
\providecommand \BibitemOpen [0]{}%
\providecommand \bibitemStop [0]{}%
\providecommand \bibitemNoStop [0]{.\EOS\space}%
\providecommand \EOS [0]{\spacefactor3000\relax}%
\providecommand \BibitemShut  [1]{\csname bibitem#1\endcsname}%
\let\auto@bib@innerbib\@empty
\bibitem [{\citenamefont {Riess}\ \emph {et~al.}(1998)\citenamefont {Riess},
  \citenamefont {Filippenko}, \citenamefont {Challis}, \citenamefont
  {Clocchiatti}, \citenamefont {Diercks} \emph {et~al.}}]{Riess:1998}%
  \BibitemOpen
  \bibfield  {author} {\bibinfo {author} {\bibfnamefont {A.~G.}\ \bibnamefont
  {Riess}}, \bibinfo {author} {\bibfnamefont {A.~V.}\ \bibnamefont
  {Filippenko}}, \bibinfo {author} {\bibfnamefont {P.}~\bibnamefont {Challis}},
  \bibinfo {author} {\bibfnamefont {A.}~\bibnamefont {Clocchiatti}}, \bibinfo
  {author} {\bibfnamefont {A.}~\bibnamefont {Diercks}},  \emph {et~al.},\
  }\href {http://stacks.iop.org/1538-3881/116/i=3/a=1009} {\bibfield  {journal}
  {\bibinfo  {journal} {The Astronomical Journal}\ }\textbf {\bibinfo {volume}
  {116}},\ \bibinfo {pages} {1009} (\bibinfo {year} {1998})}\BibitemShut
  {NoStop}%
\bibitem [{\citenamefont {Perlmutter}\ \emph {et~al.}(1999)\citenamefont
  {Perlmutter}, \citenamefont {Aldering}, \citenamefont {Goldhaber},
  \citenamefont {Knop}, \citenamefont {Nugent}, \citenamefont {others},\ and\
  \citenamefont {Project}}]{Perlmutter}%
  \BibitemOpen
  \bibfield  {author} {\bibinfo {author} {\bibfnamefont {S.}~\bibnamefont
  {Perlmutter}}, \bibinfo {author} {\bibfnamefont {G.}~\bibnamefont
  {Aldering}}, \bibinfo {author} {\bibfnamefont {G.}~\bibnamefont {Goldhaber}},
  \bibinfo {author} {\bibfnamefont {R.~A.}\ \bibnamefont {Knop}}, \bibinfo
  {author} {\bibfnamefont {P.}~\bibnamefont {Nugent}}, \bibinfo {author}
  {\bibnamefont {others}}, \ and\ \bibinfo {author} {\bibfnamefont {T.~S.~C.}\
  \bibnamefont {Project}},\ }\href
  {http://stacks.iop.org/0004-637X/517/i=2/a=565} {\bibfield  {journal}
  {\bibinfo  {journal} {The Astrophysical Journal}\ }\textbf {\bibinfo {volume}
  {517}},\ \bibinfo {pages} {565} (\bibinfo {year} {1999})}\BibitemShut
  {NoStop}%
\bibitem [{\citenamefont {Scolnic}\ and\ \citenamefont {{\it et.
  al.}}(2018)}]{Scolnic:2017caz}%
  \BibitemOpen
  \bibfield  {author} {\bibinfo {author} {\bibfnamefont {D.~M.}\ \bibnamefont
  {Scolnic}}\ and\ \bibinfo {author} {\bibnamefont {{\it et. al.}}},\ }\href
  {http://stacks.iop.org/0004-637X/859/i=2/a=101} {\bibfield  {journal}
  {\bibinfo  {journal} {The Astrophysical Journal}\ }\textbf {\bibinfo {volume}
  {859}},\ \bibinfo {pages} {101} (\bibinfo {year} {2018})}\BibitemShut
  {NoStop}%
\bibitem [{\citenamefont {Aghanim}\ \emph {et~al.}(2018)\citenamefont {Aghanim}
  \emph {et~al.}}]{Planck:2018}%
  \BibitemOpen
  \bibfield  {author} {\bibinfo {author} {\bibfnamefont {N.}~\bibnamefont
  {Aghanim}} \emph {et~al.} (\bibinfo {collaboration} {Planck}),\ }\href@noop
  {} {\  (\bibinfo {year} {2018})},\ \Eprint {http://arxiv.org/abs/1807.06209}
  {arXiv:1807.06209 [astro-ph.CO]} \BibitemShut {NoStop}%
\bibitem [{\citenamefont {Maga\~na}\ \emph {et~al.}(2018)\citenamefont
  {Maga\~na}, \citenamefont {Amante}, \citenamefont {Garc\'ia-Aspeitia},\ and\
  \citenamefont {Motta}}]{Magana:2017nfs}%
  \BibitemOpen
  \bibfield  {author} {\bibinfo {author} {\bibfnamefont {J.}~\bibnamefont
  {Maga\~na}}, \bibinfo {author} {\bibfnamefont {M.~H.}\ \bibnamefont
  {Amante}}, \bibinfo {author} {\bibfnamefont {M.~A.}\ \bibnamefont
  {Garc\'ia-Aspeitia}}, \ and\ \bibinfo {author} {\bibfnamefont
  {V.}~\bibnamefont {Motta}},\ }\href {\doibase 10.1093/mnras/sty260}
  {\bibfield  {journal} {\bibinfo  {journal} {Mon. Not. Roy. Astron. Soc.}\
  }\textbf {\bibinfo {volume} {476}},\ \bibinfo {pages} {1036} (\bibinfo {year}
  {2018})},\ \Eprint {http://arxiv.org/abs/1706.09848} {arXiv:1706.09848
  [astro-ph.CO]} \BibitemShut {NoStop}%
\bibitem [{\citenamefont {Amante}\ \emph {et~al.}(2019)\citenamefont {Amante},
  \citenamefont {Maga\~na}, \citenamefont {Motta}, \citenamefont
  {Garc\'ia-Aspeitia},\ and\ \citenamefont {Verdugo}}]{Amante:2019xao}%
  \BibitemOpen
  \bibfield  {author} {\bibinfo {author} {\bibfnamefont {M.~H.}\ \bibnamefont
  {Amante}}, \bibinfo {author} {\bibfnamefont {J.}~\bibnamefont {Maga\~na}},
  \bibinfo {author} {\bibfnamefont {V.}~\bibnamefont {Motta}}, \bibinfo
  {author} {\bibfnamefont {M.~A.}\ \bibnamefont {Garc\'ia-Aspeitia}}, \ and\
  \bibinfo {author} {\bibfnamefont {T.}~\bibnamefont {Verdugo}},\ }\href@noop
  {} {\  (\bibinfo {year} {2019})},\ \Eprint {http://arxiv.org/abs/1906.04107}
  {arXiv:1906.04107 [astro-ph.CO]} \BibitemShut {NoStop}%
\bibitem [{\citenamefont {Copeland}\ \emph {et~al.}(2006)\citenamefont
  {Copeland}, \citenamefont {Sami},\ and\ \citenamefont
  {Tsujikawa}}]{Copeland:2006wr}%
  \BibitemOpen
  \bibfield  {author} {\bibinfo {author} {\bibfnamefont {E.~J.}\ \bibnamefont
  {Copeland}}, \bibinfo {author} {\bibfnamefont {M.}~\bibnamefont {Sami}}, \
  and\ \bibinfo {author} {\bibfnamefont {S.}~\bibnamefont {Tsujikawa}},\ }\href
  {\doibase 10.1142/S021827180600942X} {\bibfield  {journal} {\bibinfo
  {journal} {Int. J. Mod. Phys.}\ }\textbf {\bibinfo {volume} {D15}},\ \bibinfo
  {pages} {1753} (\bibinfo {year} {2006})},\ \Eprint
  {http://arxiv.org/abs/hep-th/0603057} {arXiv:hep-th/0603057 [hep-th]}
  \BibitemShut {NoStop}%
\bibitem [{\citenamefont {Li}\ \emph {et~al.}(2011)\citenamefont {Li},
  \citenamefont {Li}, \citenamefont {Wang},\ and\ \citenamefont
  {Wang}}]{Li:2011sd}%
  \BibitemOpen
  \bibfield  {author} {\bibinfo {author} {\bibfnamefont {M.}~\bibnamefont
  {Li}}, \bibinfo {author} {\bibfnamefont {X.-D.}\ \bibnamefont {Li}}, \bibinfo
  {author} {\bibfnamefont {S.}~\bibnamefont {Wang}}, \ and\ \bibinfo {author}
  {\bibfnamefont {Y.}~\bibnamefont {Wang}},\ }\href {\doibase
  10.1088/0253-6102/56/3/24} {\bibfield  {journal} {\bibinfo  {journal}
  {Commun. Theor. Phys.}\ }\textbf {\bibinfo {volume} {56}},\ \bibinfo {pages}
  {525} (\bibinfo {year} {2011})},\ \Eprint {http://arxiv.org/abs/1103.5870}
  {arXiv:1103.5870 [astro-ph.CO]} \BibitemShut {NoStop}%
\bibitem [{\citenamefont {Lovelock}(1971)}]{Lovelock}%
  \BibitemOpen
  \bibfield  {author} {\bibinfo {author} {\bibfnamefont {D.}~\bibnamefont
  {Lovelock}},\ }\href {\doibase 10.1063/1.1665613} {\bibfield  {journal}
  {\bibinfo  {journal} {Journal of Mathematical Physics}\ }\textbf {\bibinfo
  {volume} {12}},\ \bibinfo {pages} {498} (\bibinfo {year} {1971})},\ \Eprint
  {http://arxiv.org/abs/https://doi.org/10.1063/1.1665613}
  {https://doi.org/10.1063/1.1665613} \BibitemShut {NoStop}%
\bibitem [{\citenamefont {Weinberg}(1989)}]{Weinberg}%
  \BibitemOpen
  \bibfield  {author} {\bibinfo {author} {\bibfnamefont {S.}~\bibnamefont
  {Weinberg}},\ }\href@noop {} {\bibfield  {journal} {\bibinfo  {journal}
  {Reviews of Modern Physics}\ }\textbf {\bibinfo {volume} {61}} (\bibinfo
  {year} {1989})}\BibitemShut {NoStop}%
\bibitem [{\citenamefont {Zeldovich}(1968)}]{Zeldovich}%
  \BibitemOpen
  \bibfield  {author} {\bibinfo {author} {\bibfnamefont {Y.~B.}\ \bibnamefont
  {Zeldovich}},\ }\href@noop {} {\bibfield  {journal} {\bibinfo  {journal}
  {Soviet Physics Uspekhi}\ }\textbf {\bibinfo {volume} {11}} (\bibinfo {year}
  {1968})}\BibitemShut {NoStop}%
\bibitem [{\citenamefont {Anderson}\ and\ \citenamefont
  {Finkelstein}(1971)}]{James}%
  \BibitemOpen
  \bibfield  {author} {\bibinfo {author} {\bibfnamefont {J.~L.}\ \bibnamefont
  {Anderson}}\ and\ \bibinfo {author} {\bibfnamefont {D.}~\bibnamefont
  {Finkelstein}},\ }\href {\doibase 10.1119/1.1986321} {\bibfield  {journal}
  {\bibinfo  {journal} {American Journal of Physics}\ }\textbf {\bibinfo
  {volume} {39}},\ \bibinfo {pages} {901} (\bibinfo {year} {1971})},\ \Eprint
  {http://arxiv.org/abs/https://doi.org/10.1119/1.1986321}
  {https://doi.org/10.1119/1.1986321} \BibitemShut {NoStop}%
\bibitem [{\citenamefont {Ellis}\ \emph {et~al.}(2011)\citenamefont {Ellis},
  \citenamefont {van Elst}, \citenamefont {Murugan},\ and\ \citenamefont
  {Uzan}}]{Ellis}%
  \BibitemOpen
  \bibfield  {author} {\bibinfo {author} {\bibfnamefont {G.~F.~R.}\
  \bibnamefont {Ellis}}, \bibinfo {author} {\bibfnamefont {H.}~\bibnamefont
  {van Elst}}, \bibinfo {author} {\bibfnamefont {J.}~\bibnamefont {Murugan}}, \
  and\ \bibinfo {author} {\bibfnamefont {J.-P.}\ \bibnamefont {Uzan}},\ }\href
  {http://stacks.iop.org/0264-9381/28/i=22/a=225007} {\bibfield  {journal}
  {\bibinfo  {journal} {Classical and Quantum Gravity}\ }\textbf {\bibinfo
  {volume} {28}},\ \bibinfo {pages} {225007} (\bibinfo {year}
  {2011})}\BibitemShut {NoStop}%
\bibitem [{\citenamefont {Gao}\ \emph {et~al.}(2014)\citenamefont {Gao},
  \citenamefont {Brandenberger}, \citenamefont {Cai},\ and\ \citenamefont
  {Chen}}]{Gao:2014nia}%
  \BibitemOpen
  \bibfield  {author} {\bibinfo {author} {\bibfnamefont {C.}~\bibnamefont
  {Gao}}, \bibinfo {author} {\bibfnamefont {R.~H.}\ \bibnamefont
  {Brandenberger}}, \bibinfo {author} {\bibfnamefont {Y.}~\bibnamefont {Cai}},
  \ and\ \bibinfo {author} {\bibfnamefont {P.}~\bibnamefont {Chen}},\ }\href
  {\doibase 10.1088/1475-7516/2014/09/021} {\bibfield  {journal} {\bibinfo
  {journal} {JCAP}\ }\textbf {\bibinfo {volume} {1409}},\ \bibinfo {pages}
  {021} (\bibinfo {year} {2014})},\ \Eprint {http://arxiv.org/abs/1405.1644}
  {arXiv:1405.1644 [gr-qc]} \BibitemShut {NoStop}%
\bibitem [{\citenamefont {Josset}\ \emph {et~al.}(2017)\citenamefont {Josset},
  \citenamefont {Perez},\ and\ \citenamefont {Sudarsky}}]{Josset:2016}%
  \BibitemOpen
  \bibfield  {author} {\bibinfo {author} {\bibfnamefont {T.}~\bibnamefont
  {Josset}}, \bibinfo {author} {\bibfnamefont {A.}~\bibnamefont {Perez}}, \
  and\ \bibinfo {author} {\bibfnamefont {D.}~\bibnamefont {Sudarsky}},\ }\href
  {\doibase 10.1103/PhysRevLett.118.021102} {\bibfield  {journal} {\bibinfo
  {journal} {Phys. Rev. Lett.}\ }\textbf {\bibinfo {volume} {118}},\ \bibinfo
  {pages} {021102} (\bibinfo {year} {2017})},\ \Eprint
  {http://arxiv.org/abs/1604.04183} {arXiv:1604.04183 [gr-qc]} \BibitemShut
  {NoStop}%
\bibitem [{\citenamefont {Perez}\ \emph {et~al.}(2018)\citenamefont {Perez},
  \citenamefont {Sudarsky},\ and\ \citenamefont {Bjorken}}]{Perez:2018wlo}%
  \BibitemOpen
  \bibfield  {author} {\bibinfo {author} {\bibfnamefont {A.}~\bibnamefont
  {Perez}}, \bibinfo {author} {\bibfnamefont {D.}~\bibnamefont {Sudarsky}}, \
  and\ \bibinfo {author} {\bibfnamefont {J.~D.}\ \bibnamefont {Bjorken}},\
  }\href {\doibase 10.1142/S0218271818460021} {\bibfield  {journal} {\bibinfo
  {journal} {Int. J. Mod. Phys.}\ }\textbf {\bibinfo {volume} {D27}},\ \bibinfo
  {pages} {1846002} (\bibinfo {year} {2018})},\ \Eprint
  {http://arxiv.org/abs/1804.07162} {arXiv:1804.07162 [gr-qc]} \BibitemShut
  {NoStop}%
\bibitem [{\citenamefont {Perez}\ and\ \citenamefont
  {Sudarsky}(2019)}]{Perez:2017krv}%
  \BibitemOpen
  \bibfield  {author} {\bibinfo {author} {\bibfnamefont {A.}~\bibnamefont
  {Perez}}\ and\ \bibinfo {author} {\bibfnamefont {D.}~\bibnamefont
  {Sudarsky}},\ }\href {\doibase 10.1103/PhysRevLett.122.221302} {\bibfield
  {journal} {\bibinfo  {journal} {Phys. Rev. Lett.}\ }\textbf {\bibinfo
  {volume} {122}},\ \bibinfo {pages} {221302} (\bibinfo {year} {2019})},\
  \Eprint {http://arxiv.org/abs/1711.05183} {arXiv:1711.05183 [gr-qc]}
  \BibitemShut {NoStop}%
\bibitem [{\citenamefont {Henneaux}\ and\ \citenamefont
  {Teitelboim}(1989)}]{Henn}%
  \BibitemOpen
  \bibfield  {author} {\bibinfo {author} {\bibfnamefont {M.}~\bibnamefont
  {Henneaux}}\ and\ \bibinfo {author} {\bibfnamefont {C.}~\bibnamefont
  {Teitelboim}},\ }\href {\doibase
  https://doi.org/10.1016/0370-2693(89)91251-3} {\bibfield  {journal} {\bibinfo
   {journal} {Physics Letters B}\ }\textbf {\bibinfo {volume} {222}},\ \bibinfo
  {pages} {195 } (\bibinfo {year} {1989})}\BibitemShut {NoStop}%
\bibitem [{\citenamefont {Abbott}\ \emph
  {et~al.}(2016{\natexlab{a}})\citenamefont {Abbott} \emph
  {et~al.}}]{Abbott:2016blz}%
  \BibitemOpen
  \bibfield  {author} {\bibinfo {author} {\bibfnamefont {B.~P.}\ \bibnamefont
  {Abbott}} \emph {et~al.} (\bibinfo {collaboration} {LIGO Scientific,
  Virgo}),\ }\href {\doibase 10.1103/PhysRevLett.116.061102} {\bibfield
  {journal} {\bibinfo  {journal} {Phys. Rev. Lett.}\ }\textbf {\bibinfo
  {volume} {116}},\ \bibinfo {pages} {061102} (\bibinfo {year}
  {2016}{\natexlab{a}})},\ \Eprint {http://arxiv.org/abs/1602.03837}
  {arXiv:1602.03837 [gr-qc]} \BibitemShut {NoStop}%
\bibitem [{\citenamefont {Abbott}\ \emph
  {et~al.}(2016{\natexlab{b}})\citenamefont {Abbott} \emph
  {et~al.}}]{Abbott:2016nmj}%
  \BibitemOpen
  \bibfield  {author} {\bibinfo {author} {\bibfnamefont {B.~P.}\ \bibnamefont
  {Abbott}} \emph {et~al.} (\bibinfo {collaboration} {LIGO Scientific,
  Virgo}),\ }\href {\doibase 10.1103/PhysRevLett.116.241103} {\bibfield
  {journal} {\bibinfo  {journal} {Phys. Rev. Lett.}\ }\textbf {\bibinfo
  {volume} {116}},\ \bibinfo {pages} {241103} (\bibinfo {year}
  {2016}{\natexlab{b}})},\ \Eprint {http://arxiv.org/abs/1606.04855}
  {arXiv:1606.04855 [gr-qc]} \BibitemShut {NoStop}%
\bibitem [{\citenamefont {et~al.}(2019)}]{MassiveBlackHoles}%
  \BibitemOpen
  \bibfield  {author} {\bibinfo {author} {\bibfnamefont {T.~E.~C.}\
  \bibnamefont {et~al.}},\ }\href
  {https://iopscience.iop.org/article/10.3847/2041-8213/ab0ec7} {\bibfield
  {journal} {\bibinfo  {journal} {ApJL}\ }\textbf {\bibinfo {volume} {875}},\
  \bibinfo {pages} {1} (\bibinfo {year} {2019})}\BibitemShut {NoStop}%
\bibitem [{\citenamefont {Bufalo}\ \emph {et~al.}(2015)\citenamefont {Bufalo},
  \citenamefont {Oksanen},\ and\ \citenamefont {Tureanu}}]{Bufalo:2015wda}%
  \BibitemOpen
  \bibfield  {author} {\bibinfo {author} {\bibfnamefont {R.}~\bibnamefont
  {Bufalo}}, \bibinfo {author} {\bibfnamefont {M.}~\bibnamefont {Oksanen}}, \
  and\ \bibinfo {author} {\bibfnamefont {A.}~\bibnamefont {Tureanu}},\ }\href
  {\doibase 10.1140/epjc/s10052-015-3683-3} {\bibfield  {journal} {\bibinfo
  {journal} {Eur. Phys. J.}\ }\textbf {\bibinfo {volume} {C75}},\ \bibinfo
  {pages} {477} (\bibinfo {year} {2015})},\ \Eprint
  {http://arxiv.org/abs/1505.04978} {arXiv:1505.04978 [hep-th]} \BibitemShut
  {NoStop}%
\bibitem [{\citenamefont {Percacci}(2018)}]{Percacci:2017fsy}%
  \BibitemOpen
  \bibfield  {author} {\bibinfo {author} {\bibfnamefont {R.}~\bibnamefont
  {Percacci}},\ }\href {\doibase 10.1007/s10701-018-0189-5} {\bibfield
  {journal} {\bibinfo  {journal} {Found. Phys.}\ }\textbf {\bibinfo {volume}
  {48}},\ \bibinfo {pages} {1364} (\bibinfo {year} {2018})},\ \Eprint
  {http://arxiv.org/abs/1712.09903} {arXiv:1712.09903 [gr-qc]} \BibitemShut
  {NoStop}%
\bibitem [{\citenamefont {Astorga-Moreno}\ \emph {et~al.}(2019)\citenamefont
  {Astorga-Moreno}, \citenamefont {Chagoya}, \citenamefont {Flores-Urbina},\
  and\ \citenamefont {Garc\'ia-Aspeitia}}]{Astorga-Moreno:2019}%
  \BibitemOpen
  \bibfield  {author} {\bibinfo {author} {\bibfnamefont {J.~A.}\ \bibnamefont
  {Astorga-Moreno}}, \bibinfo {author} {\bibfnamefont {J.}~\bibnamefont
  {Chagoya}}, \bibinfo {author} {\bibfnamefont {J.~C.}\ \bibnamefont
  {Flores-Urbina}}, \ and\ \bibinfo {author} {\bibfnamefont {M.~A.}\
  \bibnamefont {Garc\'ia-Aspeitia}},\ }\href {\doibase
  10.1088/1475-7516/2019/09/005} {\bibfield  {journal} {\bibinfo  {journal}
  {JCAP}\ }\textbf {\bibinfo {volume} {2019}},\ \bibinfo {pages} {005}
  (\bibinfo {year} {2019})},\ \Eprint {http://arxiv.org/abs/1905.11253}
  {arXiv:1905.11253 [gr-qc]} \BibitemShut {NoStop}%
\bibitem [{\citenamefont {Garc\'ia-Aspeitia}\ \emph {et~al.}(2019)\citenamefont
  {Garc\'ia-Aspeitia}, \citenamefont {Mart\'inez-Robles}, \citenamefont
  {Hern\'andez-Almada}, \citenamefont {Maga\~na},\ and\ \citenamefont
  {Motta}}]{Garcia-Aspeitia:2019yni}%
  \BibitemOpen
  \bibfield  {author} {\bibinfo {author} {\bibfnamefont {M.~A.}\ \bibnamefont
  {Garc\'ia-Aspeitia}}, \bibinfo {author} {\bibfnamefont {C.}~\bibnamefont
  {Mart\'inez-Robles}}, \bibinfo {author} {\bibfnamefont {A.}~\bibnamefont
  {Hern\'andez-Almada}}, \bibinfo {author} {\bibfnamefont {J.}~\bibnamefont
  {Maga\~na}}, \ and\ \bibinfo {author} {\bibfnamefont {V.}~\bibnamefont
  {Motta}},\ }\href {\doibase 10.1103/PhysRevD.99.123525} {\bibfield  {journal}
  {\bibinfo  {journal} {Phys. Rev.}\ }\textbf {\bibinfo {volume} {D99}},\
  \bibinfo {pages} {123525} (\bibinfo {year} {2019})},\ \Eprint
  {http://arxiv.org/abs/1903.06344} {arXiv:1903.06344 [gr-qc]} \BibitemShut
  {NoStop}%
\bibitem [{\citenamefont {Hern\'andez-Almada}\ \emph
  {et~al.}(2020)\citenamefont {Hern\'andez-Almada}, \citenamefont {Leon},
  \citenamefont {Maga\~na}, \citenamefont {Garc\'\i{}a-Aspeitia},\ and\
  \citenamefont {Motta}}]{Hernandez-Ganly:2020}%
  \BibitemOpen
  \bibfield  {author} {\bibinfo {author} {\bibfnamefont {A.}~\bibnamefont
  {Hern\'andez-Almada}}, \bibinfo {author} {\bibfnamefont {G.}~\bibnamefont
  {Leon}}, \bibinfo {author} {\bibfnamefont {J.}~\bibnamefont {Maga\~na}},
  \bibinfo {author} {\bibfnamefont {M.~A.}\ \bibnamefont
  {Garc\'\i{}a-Aspeitia}}, \ and\ \bibinfo {author} {\bibfnamefont
  {V.}~\bibnamefont {Motta}},\ }\href {\doibase 10.1093/mnras/staa2052}
  {\bibfield  {journal} {\bibinfo  {journal} {Mon. Not. Roy. Astron. Soc.}\
  }\textbf {\bibinfo {volume} {497}},\ \bibinfo {pages} {1590} (\bibinfo {year}
  {2020})},\ \Eprint {http://arxiv.org/abs/2002.12881} {arXiv:2002.12881
  [astro-ph.CO]} \BibitemShut {NoStop}%
\bibitem [{\citenamefont {Verde}\ \emph {et~al.}(2019)\citenamefont {Verde},
  \citenamefont {Treu},\ and\ \citenamefont {Riess}}]{Verde:2019ivm}%
  \BibitemOpen
  \bibfield  {author} {\bibinfo {author} {\bibfnamefont {L.}~\bibnamefont
  {Verde}}, \bibinfo {author} {\bibfnamefont {T.}~\bibnamefont {Treu}}, \ and\
  \bibinfo {author} {\bibfnamefont {A.~G.}\ \bibnamefont {Riess}}\ }(\bibinfo
  {year} {2019})\ \Eprint {http://arxiv.org/abs/1907.10625} {arXiv:1907.10625
  [astro-ph.CO]} \BibitemShut {NoStop}%
\bibitem [{\citenamefont {Perez}\ \emph {et~al.}(2020)\citenamefont {Perez},
  \citenamefont {Sudarsky},\ and\ \citenamefont
  {Wilson-Ewing}}]{Perez:2020cwa}%
  \BibitemOpen
  \bibfield  {author} {\bibinfo {author} {\bibfnamefont {A.}~\bibnamefont
  {Perez}}, \bibinfo {author} {\bibfnamefont {D.}~\bibnamefont {Sudarsky}}, \
  and\ \bibinfo {author} {\bibfnamefont {E.}~\bibnamefont {Wilson-Ewing}},\
  }\href@noop {} {\  (\bibinfo {year} {2020})},\ \Eprint
  {http://arxiv.org/abs/2001.07536} {arXiv:2001.07536 [astro-ph.CO]}
  \BibitemShut {NoStop}%
\bibitem [{\citenamefont {Di~Valentino}\ \emph {et~al.}(2021)\citenamefont
  {Di~Valentino}, \citenamefont {Mena}, \citenamefont {Pan}, \citenamefont
  {Visinelli}, \citenamefont {Yang}, \citenamefont {Melchiorri}, \citenamefont
  {Mota}, \citenamefont {Riess},\ and\ \citenamefont
  {Silk}}]{DiValentino:2021izs}%
  \BibitemOpen
  \bibfield  {author} {\bibinfo {author} {\bibfnamefont {E.}~\bibnamefont
  {Di~Valentino}}, \bibinfo {author} {\bibfnamefont {O.}~\bibnamefont {Mena}},
  \bibinfo {author} {\bibfnamefont {S.}~\bibnamefont {Pan}}, \bibinfo {author}
  {\bibfnamefont {L.}~\bibnamefont {Visinelli}}, \bibinfo {author}
  {\bibfnamefont {W.}~\bibnamefont {Yang}}, \bibinfo {author} {\bibfnamefont
  {A.}~\bibnamefont {Melchiorri}}, \bibinfo {author} {\bibfnamefont {D.~F.}\
  \bibnamefont {Mota}}, \bibinfo {author} {\bibfnamefont {A.~G.}\ \bibnamefont
  {Riess}}, \ and\ \bibinfo {author} {\bibfnamefont {J.}~\bibnamefont {Silk}},\
  }\href@noop {} {\  (\bibinfo {year} {2021})},\ \Eprint
  {http://arxiv.org/abs/2103.01183} {arXiv:2103.01183 [astro-ph.CO]}
  \BibitemShut {NoStop}%
\bibitem [{\citenamefont {Efstathiou}(2021)}]{Efstathiou:2021ocp}%
  \BibitemOpen
  \bibfield  {author} {\bibinfo {author} {\bibfnamefont {G.}~\bibnamefont
  {Efstathiou}},\ }\href@noop {} {\  (\bibinfo {year} {2021})},\ \Eprint
  {http://arxiv.org/abs/2103.08723} {arXiv:2103.08723 [astro-ph.CO]}
  \BibitemShut {NoStop}%
\bibitem [{\citenamefont {Basak}\ \emph {et~al.}(2016)\citenamefont {Basak},
  \citenamefont {Fabre},\ and\ \citenamefont {Shankaranarayanan}}]{Basak:2015}%
  \BibitemOpen
  \bibfield  {author} {\bibinfo {author} {\bibfnamefont {A.}~\bibnamefont
  {Basak}}, \bibinfo {author} {\bibfnamefont {O.}~\bibnamefont {Fabre}}, \ and\
  \bibinfo {author} {\bibfnamefont {S.}~\bibnamefont {Shankaranarayanan}},\
  }\href {\doibase 10.1007/s10714-016-2116-4} {\bibfield  {journal} {\bibinfo
  {journal} {Gen. Rel. Grav.}\ }\textbf {\bibinfo {volume} {48}},\ \bibinfo
  {pages} {123} (\bibinfo {year} {2016})},\ \Eprint
  {http://arxiv.org/abs/1511.01805} {arXiv:1511.01805 [gr-qc]} \BibitemShut
  {NoStop}%
\bibitem [{\citenamefont {Alvarez}\ \emph {et~al.}(2015)\citenamefont
  {Alvarez}, \citenamefont {Gonz\'alez-Mart\'in}, \citenamefont
  {Herrero-Valea},\ and\ \citenamefont {Mart\'in}}]{Alvarez:2015sba}%
  \BibitemOpen
  \bibfield  {author} {\bibinfo {author} {\bibfnamefont {E.}~\bibnamefont
  {Alvarez}}, \bibinfo {author} {\bibfnamefont {S.}~\bibnamefont
  {Gonz\'alez-Mart\'in}}, \bibinfo {author} {\bibfnamefont {M.}~\bibnamefont
  {Herrero-Valea}}, \ and\ \bibinfo {author} {\bibfnamefont {C.~P.}\
  \bibnamefont {Mart\'in}},\ }\href {\doibase 10.1007/JHEP08(2015)078}
  {\bibfield  {journal} {\bibinfo  {journal} {JHEP}\ }\textbf {\bibinfo
  {volume} {08}},\ \bibinfo {pages} {078} (\bibinfo {year} {2015})},\ \Eprint
  {http://arxiv.org/abs/1505.01995} {arXiv:1505.01995 [hep-th]} \BibitemShut
  {NoStop}%
\bibitem [{\citenamefont {Zhang}\ \emph {et~al.}(2017)\citenamefont {Zhang},
  \citenamefont {Li},\ and\ \citenamefont {Xia}}]{Zhang:2016}%
  \BibitemOpen
  \bibfield  {author} {\bibinfo {author} {\bibfnamefont {M.-J.}\ \bibnamefont
  {Zhang}}, \bibinfo {author} {\bibfnamefont {H.}~\bibnamefont {Li}}, \ and\
  \bibinfo {author} {\bibfnamefont {J.-Q.}\ \bibnamefont {Xia}},\ }\href
  {\doibase 10.1140/epjc/s10052-017-5005-4} {\bibfield  {journal} {\bibinfo
  {journal} {Eur. Phys. J.}\ }\textbf {\bibinfo {volume} {C77}},\ \bibinfo
  {pages} {434} (\bibinfo {year} {2017})},\ \Eprint
  {http://arxiv.org/abs/1601.01758} {arXiv:1601.01758 [astro-ph.CO]}
  \BibitemShut {NoStop}%
\bibitem [{\citenamefont {Al~Mamon}\ and\ \citenamefont
  {Bamba}(2018)}]{Mamon:2018dxf}%
  \BibitemOpen
  \bibfield  {author} {\bibinfo {author} {\bibfnamefont {A.}~\bibnamefont
  {Al~Mamon}}\ and\ \bibinfo {author} {\bibfnamefont {K.}~\bibnamefont
  {Bamba}},\ }\href {\doibase 10.1140/epjc/s10052-018-6355-2} {\bibfield
  {journal} {\bibinfo  {journal} {Eur. Phys. J.}\ }\textbf {\bibinfo {volume}
  {C78}},\ \bibinfo {pages} {862} (\bibinfo {year} {2018})},\ \Eprint
  {http://arxiv.org/abs/1805.02854} {arXiv:1805.02854 [gr-qc]} \BibitemShut
  {NoStop}%
\bibitem [{\citenamefont {Velten}\ and\ \citenamefont
  {Caram\^es}(2021)}]{Velten:2021xxw}%
  \BibitemOpen
  \bibfield  {author} {\bibinfo {author} {\bibfnamefont {H.}~\bibnamefont
  {Velten}}\ and\ \bibinfo {author} {\bibfnamefont {T.~R.~P.}\ \bibnamefont
  {Caram\^es}},\ }\href {\doibase 10.3390/universe7020038} {\bibfield
  {journal} {\bibinfo  {journal} {Universe}\ }\textbf {\bibinfo {volume} {7}},\
  \bibinfo {pages} {38} (\bibinfo {year} {2021})},\ \Eprint
  {http://arxiv.org/abs/2102.03457} {arXiv:2102.03457 [gr-qc]} \BibitemShut
  {NoStop}%
\bibitem [{\citenamefont {Ballentine}(1991)}]{Ballentine}%
  \BibitemOpen
  \bibfield  {author} {\bibinfo {author} {\bibfnamefont {L.~E.}\ \bibnamefont
  {Ballentine}},\ }\href {\doibase 10.1103/PhysRevA.43.9} {\bibfield  {journal}
  {\bibinfo  {journal} {Phys. Rev. A}\ }\textbf {\bibinfo {volume} {43}},\
  \bibinfo {pages} {9} (\bibinfo {year} {1991})}\BibitemShut {NoStop}%
\bibitem [{\citenamefont {Bassi}\ \emph {et~al.}(2005)\citenamefont {Bassi},
  \citenamefont {Ippoliti},\ and\ \citenamefont {Vacchini}}]{Bassi_2005}%
  \BibitemOpen
  \bibfield  {author} {\bibinfo {author} {\bibfnamefont {A.}~\bibnamefont
  {Bassi}}, \bibinfo {author} {\bibfnamefont {E.}~\bibnamefont {Ippoliti}}, \
  and\ \bibinfo {author} {\bibfnamefont {B.}~\bibnamefont {Vacchini}},\ }\href
  {\doibase 10.1088/0305-4470/38/37/007} {\bibfield  {journal} {\bibinfo
  {journal} {Journal of Physics A: Mathematical and General}\ }\textbf
  {\bibinfo {volume} {38}},\ \bibinfo {pages} {8017} (\bibinfo {year}
  {2005})}\BibitemShut {NoStop}%
\bibitem [{\citenamefont {Corral}\ \emph {et~al.}(2020)\citenamefont {Corral},
  \citenamefont {Cruz},\ and\ \citenamefont {González}}]{Corral:2020}%
  \BibitemOpen
  \bibfield  {author} {\bibinfo {author} {\bibfnamefont {C.}~\bibnamefont
  {Corral}}, \bibinfo {author} {\bibfnamefont {N.}~\bibnamefont {Cruz}}, \ and\
  \bibinfo {author} {\bibfnamefont {E.}~\bibnamefont {González}},\ }\href
  {\doibase 10.1103/PhysRevD.102.023508} {\bibfield  {journal} {\bibinfo
  {journal} {Phys. Rev. D}\ }\textbf {\bibinfo {volume} {102}},\ \bibinfo
  {pages} {023508} (\bibinfo {year} {2020})},\ \Eprint
  {http://arxiv.org/abs/2005.06052} {arXiv:2005.06052 [gr-qc]} \BibitemShut
  {NoStop}%
\bibitem [{\citenamefont {Komatsu}\ \emph {et~al.}(2011)\citenamefont
  {Komatsu}, \citenamefont {Smith}, \citenamefont {Dunkley}, \citenamefont
  {Bennett}, \citenamefont {Gold}, \citenamefont {Hinshaw}, \citenamefont
  {Jarosik}, \citenamefont {Larson}, \citenamefont {Nolta}, \citenamefont
  {Page}, \citenamefont {Spergel}, \citenamefont {Halpern}, \citenamefont
  {Hill}, \citenamefont {Kogut}, \citenamefont {Limon}, \citenamefont {Meyer},
  \citenamefont {Odegard}, \citenamefont {Tucker}, \citenamefont {Weiland},
  \citenamefont {Wollack},\ and\ \citenamefont {Wright}}]{Komatsu:2011}%
  \BibitemOpen
  \bibfield  {author} {\bibinfo {author} {\bibfnamefont {E.}~\bibnamefont
  {Komatsu}}, \bibinfo {author} {\bibfnamefont {K.~M.}\ \bibnamefont {Smith}},
  \bibinfo {author} {\bibfnamefont {J.}~\bibnamefont {Dunkley}}, \bibinfo
  {author} {\bibfnamefont {C.~L.}\ \bibnamefont {Bennett}}, \bibinfo {author}
  {\bibfnamefont {B.}~\bibnamefont {Gold}}, \bibinfo {author} {\bibfnamefont
  {G.}~\bibnamefont {Hinshaw}}, \bibinfo {author} {\bibfnamefont
  {N.}~\bibnamefont {Jarosik}}, \bibinfo {author} {\bibfnamefont
  {D.}~\bibnamefont {Larson}}, \bibinfo {author} {\bibfnamefont {M.~R.}\
  \bibnamefont {Nolta}}, \bibinfo {author} {\bibfnamefont {L.}~\bibnamefont
  {Page}}, \bibinfo {author} {\bibfnamefont {D.~N.}\ \bibnamefont {Spergel}},
  \bibinfo {author} {\bibfnamefont {M.}~\bibnamefont {Halpern}}, \bibinfo
  {author} {\bibfnamefont {R.~S.}\ \bibnamefont {Hill}}, \bibinfo {author}
  {\bibfnamefont {A.}~\bibnamefont {Kogut}}, \bibinfo {author} {\bibfnamefont
  {M.}~\bibnamefont {Limon}}, \bibinfo {author} {\bibfnamefont {S.~S.}\
  \bibnamefont {Meyer}}, \bibinfo {author} {\bibfnamefont {N.}~\bibnamefont
  {Odegard}}, \bibinfo {author} {\bibfnamefont {G.~S.}\ \bibnamefont {Tucker}},
  \bibinfo {author} {\bibfnamefont {J.~L.}\ \bibnamefont {Weiland}}, \bibinfo
  {author} {\bibfnamefont {E.}~\bibnamefont {Wollack}}, \ and\ \bibinfo
  {author} {\bibfnamefont {E.~L.}\ \bibnamefont {Wright}},\ }\href
  {http://stacks.iop.org/0067-0049/192/i=2/a=18} {\bibfield  {journal}
  {\bibinfo  {journal} {The Astrophysical Journal Supplement Series}\ }\textbf
  {\bibinfo {volume} {192}},\ \bibinfo {pages} {18} (\bibinfo {year}
  {2011})}\BibitemShut {NoStop}%
\bibitem [{\citenamefont {Paoletti}\ \emph {et~al.}(2020)\citenamefont
  {Paoletti}, \citenamefont {Hazra}, \citenamefont {Finelli},\ and\
  \citenamefont {Smoot}}]{Paoletti:2020ndu}%
  \BibitemOpen
  \bibfield  {author} {\bibinfo {author} {\bibfnamefont {D.}~\bibnamefont
  {Paoletti}}, \bibinfo {author} {\bibfnamefont {D.~K.}\ \bibnamefont {Hazra}},
  \bibinfo {author} {\bibfnamefont {F.}~\bibnamefont {Finelli}}, \ and\
  \bibinfo {author} {\bibfnamefont {G.~F.}\ \bibnamefont {Smoot}},\ }\href
  {\doibase 10.1088/1475-7516/2020/09/005} {\bibfield  {journal} {\bibinfo
  {journal} {JCAP}\ }\textbf {\bibinfo {volume} {09}},\ \bibinfo {pages} {005}
  (\bibinfo {year} {2020})},\ \Eprint {http://arxiv.org/abs/2005.12222}
  {arXiv:2005.12222 [astro-ph.CO]} \BibitemShut {NoStop}%
\bibitem [{\citenamefont {Riess}\ \emph {et~al.}(2016)\citenamefont {Riess}
  \emph {et~al.}}]{Riess:2016jrr}%
  \BibitemOpen
  \bibfield  {author} {\bibinfo {author} {\bibfnamefont {A.~G.}\ \bibnamefont
  {Riess}} \emph {et~al.},\ }\href {\doibase 10.3847/0004-637X/826/1/56}
  {\bibfield  {journal} {\bibinfo  {journal} {Astrophys. J.}\ }\textbf
  {\bibinfo {volume} {826}},\ \bibinfo {pages} {56} (\bibinfo {year} {2016})},\
  \Eprint {http://arxiv.org/abs/1604.01424} {arXiv:1604.01424 [astro-ph.CO]}
  \BibitemShut {NoStop}%
\bibitem [{\citenamefont {Riess}\ \emph {et~al.}(2019)\citenamefont {Riess},
  \citenamefont {Casertano}, \citenamefont {Yuan}, \citenamefont {Macri},\ and\
  \citenamefont {Scolnic}}]{Riess:2019cxk}%
  \BibitemOpen
  \bibfield  {author} {\bibinfo {author} {\bibfnamefont {A.~G.}\ \bibnamefont
  {Riess}}, \bibinfo {author} {\bibfnamefont {S.}~\bibnamefont {Casertano}},
  \bibinfo {author} {\bibfnamefont {W.}~\bibnamefont {Yuan}}, \bibinfo {author}
  {\bibfnamefont {L.~M.}\ \bibnamefont {Macri}}, \ and\ \bibinfo {author}
  {\bibfnamefont {D.}~\bibnamefont {Scolnic}},\ }\href {\doibase
  10.3847/1538-4357/ab1422} {\bibfield  {journal} {\bibinfo  {journal}
  {Astrophys. J.}\ }\textbf {\bibinfo {volume} {876}},\ \bibinfo {pages} {85}
  (\bibinfo {year} {2019})},\ \Eprint {http://arxiv.org/abs/1903.07603}
  {arXiv:1903.07603 [astro-ph.CO]} \BibitemShut {NoStop}%
\bibitem [{\citenamefont {{Foreman-Mackey}}\ \emph {et~al.}(2013)\citenamefont
  {{Foreman-Mackey}}, \citenamefont {{Hogg}}, \citenamefont {{Lang}},\ and\
  \citenamefont {{Goodman}}}]{Foreman:2013}%
  \BibitemOpen
  \bibfield  {author} {\bibinfo {author} {\bibfnamefont {D.}~\bibnamefont
  {{Foreman-Mackey}}}, \bibinfo {author} {\bibfnamefont {D.~W.}\ \bibnamefont
  {{Hogg}}}, \bibinfo {author} {\bibfnamefont {D.}~\bibnamefont {{Lang}}}, \
  and\ \bibinfo {author} {\bibfnamefont {J.}~\bibnamefont {{Goodman}}},\ }\href
  {\doibase 10.1086/670067} {\bibfield  {journal} {\bibinfo  {journal} {PASP}\
  }\textbf {\bibinfo {volume} {125}},\ \bibinfo {pages} {306} (\bibinfo {year}
  {2013})},\ \Eprint {http://arxiv.org/abs/1202.3665} {arXiv:1202.3665
  [astro-ph.IM]} \BibitemShut {NoStop}%
\bibitem [{\citenamefont {Gelman}\ and\ \citenamefont
  {Rubin}(1992)}]{Gelman:1992}%
  \BibitemOpen
  \bibfield  {author} {\bibinfo {author} {\bibfnamefont {A.}~\bibnamefont
  {Gelman}}\ and\ \bibinfo {author} {\bibfnamefont {D.}~\bibnamefont {Rubin}},\
  }\href {\doibase 10.1103/PhysRevD.67.101301} {\bibfield  {journal} {\bibinfo
  {journal} {Statistical Science}\ }\textbf {\bibinfo {volume} {67}},\ \bibinfo
  {pages} {457} (\bibinfo {year} {1992})}\BibitemShut {NoStop}%
\bibitem [{\citenamefont {Conley}\ \emph {et~al.}(2010)\citenamefont {Conley},
  \citenamefont {Guy}, \citenamefont {Sullivan}, \citenamefont {Regnault},
  \citenamefont {Astier}, \citenamefont {Balland}, \citenamefont {Basa},
  \citenamefont {Carlberg}, \citenamefont {Fouchez}, \citenamefont {Hardin},
  \citenamefont {Hook}, \citenamefont {Howell}, \citenamefont {Pain},
  \citenamefont {Palanque-Delabrouille}, \citenamefont {Perrett}, \citenamefont
  {Pritchet}, \citenamefont {Rich}, \citenamefont {Ruhlmann-Kleider},
  \citenamefont {Balam}, \citenamefont {Baumont}, \citenamefont {Ellis},
  \citenamefont {Fabbro}, \citenamefont {Fakhouri}, \citenamefont {Fourmanoit},
  \citenamefont {Gonz{\'{a}}lez-Gait{\'{a}}n}, \citenamefont {Graham},
  \citenamefont {Hudson}, \citenamefont {Hsiao}, \citenamefont {Kronborg},
  \citenamefont {Lidman}, \citenamefont {Mourao}, \citenamefont {Neill},
  \citenamefont {Perlmutter}, \citenamefont {Ripoche}, \citenamefont {Suzuki},\
  and\ \citenamefont {Walker}}]{Conley_2010}%
  \BibitemOpen
  \bibfield  {author} {\bibinfo {author} {\bibfnamefont {A.}~\bibnamefont
  {Conley}}, \bibinfo {author} {\bibfnamefont {J.}~\bibnamefont {Guy}},
  \bibinfo {author} {\bibfnamefont {M.}~\bibnamefont {Sullivan}}, \bibinfo
  {author} {\bibfnamefont {N.}~\bibnamefont {Regnault}}, \bibinfo {author}
  {\bibfnamefont {P.}~\bibnamefont {Astier}}, \bibinfo {author} {\bibfnamefont
  {C.}~\bibnamefont {Balland}}, \bibinfo {author} {\bibfnamefont
  {S.}~\bibnamefont {Basa}}, \bibinfo {author} {\bibfnamefont {R.~G.}\
  \bibnamefont {Carlberg}}, \bibinfo {author} {\bibfnamefont {D.}~\bibnamefont
  {Fouchez}}, \bibinfo {author} {\bibfnamefont {D.}~\bibnamefont {Hardin}},
  \bibinfo {author} {\bibfnamefont {I.~M.}\ \bibnamefont {Hook}}, \bibinfo
  {author} {\bibfnamefont {D.~A.}\ \bibnamefont {Howell}}, \bibinfo {author}
  {\bibfnamefont {R.}~\bibnamefont {Pain}}, \bibinfo {author} {\bibfnamefont
  {N.}~\bibnamefont {Palanque-Delabrouille}}, \bibinfo {author} {\bibfnamefont
  {K.~M.}\ \bibnamefont {Perrett}}, \bibinfo {author} {\bibfnamefont {C.~J.}\
  \bibnamefont {Pritchet}}, \bibinfo {author} {\bibfnamefont {J.}~\bibnamefont
  {Rich}}, \bibinfo {author} {\bibfnamefont {V.}~\bibnamefont
  {Ruhlmann-Kleider}}, \bibinfo {author} {\bibfnamefont {D.}~\bibnamefont
  {Balam}}, \bibinfo {author} {\bibfnamefont {S.}~\bibnamefont {Baumont}},
  \bibinfo {author} {\bibfnamefont {R.~S.}\ \bibnamefont {Ellis}}, \bibinfo
  {author} {\bibfnamefont {S.}~\bibnamefont {Fabbro}}, \bibinfo {author}
  {\bibfnamefont {H.~K.}\ \bibnamefont {Fakhouri}}, \bibinfo {author}
  {\bibfnamefont {N.}~\bibnamefont {Fourmanoit}}, \bibinfo {author}
  {\bibfnamefont {S.}~\bibnamefont {Gonz{\'{a}}lez-Gait{\'{a}}n}}, \bibinfo
  {author} {\bibfnamefont {M.~L.}\ \bibnamefont {Graham}}, \bibinfo {author}
  {\bibfnamefont {M.~J.}\ \bibnamefont {Hudson}}, \bibinfo {author}
  {\bibfnamefont {E.}~\bibnamefont {Hsiao}}, \bibinfo {author} {\bibfnamefont
  {T.}~\bibnamefont {Kronborg}}, \bibinfo {author} {\bibfnamefont
  {C.}~\bibnamefont {Lidman}}, \bibinfo {author} {\bibfnamefont {A.~M.}\
  \bibnamefont {Mourao}}, \bibinfo {author} {\bibfnamefont {J.~D.}\
  \bibnamefont {Neill}}, \bibinfo {author} {\bibfnamefont {S.}~\bibnamefont
  {Perlmutter}}, \bibinfo {author} {\bibfnamefont {P.}~\bibnamefont {Ripoche}},
  \bibinfo {author} {\bibfnamefont {N.}~\bibnamefont {Suzuki}}, \ and\ \bibinfo
  {author} {\bibfnamefont {E.~S.}\ \bibnamefont {Walker}},\ }\href {\doibase
  10.1088/0067-0049/192/1/1} {\bibfield  {journal} {\bibinfo  {journal} {The
  Astrophysical Journal Supplement Series}\ }\textbf {\bibinfo {volume}
  {192}},\ \bibinfo {pages} {1} (\bibinfo {year} {2010})}\BibitemShut {NoStop}%
\bibitem [{\citenamefont {Chen}\ \emph {et~al.}(2019)\citenamefont {Chen},
  \citenamefont {Huang},\ and\ \citenamefont {Wang}}]{Chen:2018dbv}%
  \BibitemOpen
  \bibfield  {author} {\bibinfo {author} {\bibfnamefont {L.}~\bibnamefont
  {Chen}}, \bibinfo {author} {\bibfnamefont {Q.-G.}\ \bibnamefont {Huang}}, \
  and\ \bibinfo {author} {\bibfnamefont {K.}~\bibnamefont {Wang}},\ }\href
  {\doibase 10.1088/1475-7516/2019/02/028} {\bibfield  {journal} {\bibinfo
  {journal} {JCAP}\ }\textbf {\bibinfo {volume} {02}},\ \bibinfo {pages} {028}
  (\bibinfo {year} {2019})},\ \Eprint {http://arxiv.org/abs/1808.05724}
  {arXiv:1808.05724 [astro-ph.CO]} \BibitemShut {NoStop}%
\bibitem [{\citenamefont {Nunes}\ \emph {et~al.}(2020)\citenamefont {Nunes},
  \citenamefont {Yadav}, \citenamefont {Jesus},\ and\ \citenamefont
  {Bernui}}]{nunes:2020}%
  \BibitemOpen
  \bibfield  {author} {\bibinfo {author} {\bibfnamefont {R.~C.}\ \bibnamefont
  {Nunes}}, \bibinfo {author} {\bibfnamefont {S.~K.}\ \bibnamefont {Yadav}},
  \bibinfo {author} {\bibfnamefont {J.~F.}\ \bibnamefont {Jesus}}, \ and\
  \bibinfo {author} {\bibfnamefont {A.}~\bibnamefont {Bernui}},\ }\href@noop {}
  {\  (\bibinfo {year} {2020})},\ \Eprint {http://arxiv.org/abs/2002.09293}
  {arXiv:2002.09293 [astro-ph.CO]} \BibitemShut {NoStop}%
\bibitem [{\citenamefont {Bouwens}\ \emph {et~al.}(2015)\citenamefont
  {Bouwens}, \citenamefont {Illingworth}, \citenamefont {Oesch}, \citenamefont
  {Caruana}, \citenamefont {Holwerda}, \citenamefont {Smit},\ and\
  \citenamefont {Wilkins}}]{Bouwens:2015vha}%
  \BibitemOpen
  \bibfield  {author} {\bibinfo {author} {\bibfnamefont {R.~J.}\ \bibnamefont
  {Bouwens}}, \bibinfo {author} {\bibfnamefont {G.~D.}\ \bibnamefont
  {Illingworth}}, \bibinfo {author} {\bibfnamefont {P.~A.}\ \bibnamefont
  {Oesch}}, \bibinfo {author} {\bibfnamefont {J.}~\bibnamefont {Caruana}},
  \bibinfo {author} {\bibfnamefont {B.}~\bibnamefont {Holwerda}}, \bibinfo
  {author} {\bibfnamefont {R.}~\bibnamefont {Smit}}, \ and\ \bibinfo {author}
  {\bibfnamefont {S.}~\bibnamefont {Wilkins}},\ }\href {\doibase
  10.1088/0004-637X/811/2/140} {\bibfield  {journal} {\bibinfo  {journal}
  {Astrophys. J.}\ }\textbf {\bibinfo {volume} {811}},\ \bibinfo {pages} {140}
  (\bibinfo {year} {2015})},\ \Eprint {http://arxiv.org/abs/1503.08228}
  {arXiv:1503.08228 [astro-ph.CO]} \BibitemShut {NoStop}%
\bibitem [{\citenamefont {Liu}\ and\ \citenamefont {Shaw}(2020)}]{Liu:2019awk}%
  \BibitemOpen
  \bibfield  {author} {\bibinfo {author} {\bibfnamefont {A.}~\bibnamefont
  {Liu}}\ and\ \bibinfo {author} {\bibfnamefont {J.~R.}\ \bibnamefont {Shaw}},\
  }\href {\doibase 10.1088/1538-3873/ab5bfd} {\bibfield  {journal} {\bibinfo
  {journal} {Publ. Astron. Soc. Pac.}\ }\textbf {\bibinfo {volume} {132}},\
  \bibinfo {pages} {062001} (\bibinfo {year} {2020})},\ \Eprint
  {http://arxiv.org/abs/1907.08211} {arXiv:1907.08211 [astro-ph.IM]}
  \BibitemShut {NoStop}%
\bibitem [{\citenamefont {{Magg}}\ \emph {et~al.}(2019)\citenamefont {{Magg}},
  \citenamefont {{Klessen}}, \citenamefont {{Glover}},\ and\ \citenamefont
  {{Li}}}]{Magg}%
  \BibitemOpen
  \bibfield  {author} {\bibinfo {author} {\bibfnamefont {M.}~\bibnamefont
  {{Magg}}}, \bibinfo {author} {\bibfnamefont {R.~S.}\ \bibnamefont
  {{Klessen}}}, \bibinfo {author} {\bibfnamefont {S.~C.~O.}\ \bibnamefont
  {{Glover}}}, \ and\ \bibinfo {author} {\bibfnamefont {H.}~\bibnamefont
  {{Li}}},\ }\href {\doibase 10.1093/mnras/stz1210} {\bibfield  {journal}
  {\bibinfo  {journal} {MNRAS}\ }\textbf {\bibinfo {volume} {487}},\ \bibinfo
  {pages} {486} (\bibinfo {year} {2019})},\ \Eprint
  {http://arxiv.org/abs/1903.08661} {arXiv:1903.08661 [astro-ph.GA]}
  \BibitemShut {NoStop}%
\bibitem [{\citenamefont {{Glover}}(2013)}]{Glover}%
  \BibitemOpen
  \bibfield  {author} {\bibinfo {author} {\bibfnamefont {S.}~\bibnamefont
  {{Glover}}},\ }in\ \href {\doibase 10.1007/978-3-642-32362-1_3} {\emph
  {\bibinfo {booktitle} {The First Galaxies}}},\ \bibinfo {series}
  {Astrophysics and Space Science Library}, Vol.\ \bibinfo {volume} {396},\
  \bibinfo {editor} {edited by\ \bibinfo {editor} {\bibfnamefont
  {T.}~\bibnamefont {{Wiklind}}}, \bibinfo {editor} {\bibfnamefont
  {B.}~\bibnamefont {{Mobasher}}}, \ and\ \bibinfo {editor} {\bibfnamefont
  {V.}~\bibnamefont {{Bromm}}}}\ (\bibinfo {year} {2013})\ p.\ \bibinfo {pages}
  {103},\ \Eprint {http://arxiv.org/abs/1209.2509} {arXiv:1209.2509
  [astro-ph.CO]} \BibitemShut {NoStop}%
\bibitem [{\citenamefont {{Trenti}}(2010)}]{Trenti}%
  \BibitemOpen
  \bibfield  {author} {\bibinfo {author} {\bibfnamefont {M.}~\bibnamefont
  {{Trenti}}},\ }in\ \href {\doibase 10.1063/1.3518841} {\emph {\bibinfo
  {booktitle} {American Institute of Physics Conference Series}}},\ \bibinfo
  {series} {American Institute of Physics Conference Series}, Vol.\ \bibinfo
  {volume} {1294},\ \bibinfo {editor} {edited by\ \bibinfo {editor}
  {\bibfnamefont {D.~J.}\ \bibnamefont {{Whalen}}}, \bibinfo {editor}
  {\bibfnamefont {V.}~\bibnamefont {{Bromm}}}, \ and\ \bibinfo {editor}
  {\bibfnamefont {N.}~\bibnamefont {{Yoshida}}}}\ (\bibinfo {year} {2010})\
  pp.\ \bibinfo {pages} {134--137},\ \Eprint {http://arxiv.org/abs/1006.4434}
  {arXiv:1006.4434 [astro-ph.CO]} \BibitemShut {NoStop}%
\bibitem [{\citenamefont {Lombriser}(2018)}]{Lombriser:2018aru}%
  \BibitemOpen
  \bibfield  {author} {\bibinfo {author} {\bibfnamefont {L.}~\bibnamefont
  {Lombriser}},\ }\href {\doibase 10.1088/1475-7516/2019/09/065} {\  (\bibinfo
  {year} {2018}),\ 10.1088/1475-7516/2019/09/065},\ \bibinfo {note}
  {[JCAP1909,no.09,065(2019)]},\ \Eprint {http://arxiv.org/abs/1805.05918}
  {arXiv:1805.05918 [astro-ph.CO]} \BibitemShut {NoStop}%
\bibitem [{\citenamefont {Bowman}\ \emph {et~al.}(2018)\citenamefont {Bowman},
  \citenamefont {Rogers}, \citenamefont {Monsalve}, \citenamefont {Mozdzen},\
  and\ \citenamefont {Mahesh}}]{Bowman:2018yin}%
  \BibitemOpen
  \bibfield  {author} {\bibinfo {author} {\bibfnamefont {J.~D.}\ \bibnamefont
  {Bowman}}, \bibinfo {author} {\bibfnamefont {A.~E.~E.}\ \bibnamefont
  {Rogers}}, \bibinfo {author} {\bibfnamefont {R.~A.}\ \bibnamefont
  {Monsalve}}, \bibinfo {author} {\bibfnamefont {T.~J.}\ \bibnamefont
  {Mozdzen}}, \ and\ \bibinfo {author} {\bibfnamefont {N.}~\bibnamefont
  {Mahesh}},\ }\href {\doibase 10.1038/nature25792} {\bibfield  {journal}
  {\bibinfo  {journal} {Nature}\ }\textbf {\bibinfo {volume} {555}},\ \bibinfo
  {pages} {67} (\bibinfo {year} {2018})},\ \Eprint
  {http://arxiv.org/abs/1810.05912} {arXiv:1810.05912 [astro-ph.CO]}
  \BibitemShut {NoStop}%
\end{thebibliography}%


%

\end{document}